\documentclass[a4paper,11pt]{article}
\pdfoutput=1 

\usepackage{jinstpub} 

\usepackage[utf8]{inputenc}
\usepackage[T1]{fontenc}
\usepackage{subfloat}
\usepackage{subcaption}
\usepackage{caption}
\usepackage{graphicx}

\title{Study of water Cherenkov detector design for ground-based gamma-ray experiments}

\author[1]{F. Bisconti\note{Corresponding author.}}
\author{and A. Chiavassa}
\affiliation{University of Turin,\\Via Pietro Giuria 1, Turin, Italy}

\emailAdd{francesca.bisconti@to.infn.it}

\abstract{In the framework of the development of the SWGO experiment we have performed a detailed study of the single unit of an extensive air shower observatory based on an array of water Cherenkov detectors. Indeed, one of the possible water Cherenkov detector unit configurations for SWGO consists of tanks, and to reach a high detection efficiency and discrimination capability between gamma-ray and hadronic air showers, different tank designs are under investigation. In this study, we considered double-layer tanks with several sizes, shapes and number of photo-multiplier tubes (PMTs). Muons, electrons, and gamma-rays with energies typical of secondary particles in extensive air showers have been simulated entering the tanks with zenith angles from 0 to 60 degrees. The tank response was evaluated considering the number of photoelectrons produced by the PMTs, the detection efficiency, and the time resolution of the measurement of the first photon. This analysis allowed to compare the performance of tanks with different size, configuration of PMTs, and with circular, hexagonal and square geometry. The method used and the results will be discussed in this paper.}


\begin{document}
\maketitle
\flushbottom

\section{Introduction}\label{sec:intro}

Wide field of view gamma-ray observatories can be realized by an array of water Cherenkov detectors, covering areas ranging from $10^4$ to $10^6$ square meter, usually located in desertic areas. Secondary particles produced in extensive air showers induced by astrophysical gamma-rays or hadrons (that represent a background source), can be detected measuring the Cherenkov light produced when they cross the detectors filled with clean water. 
A next-generation gamma-ray observatory is the SWGO experiment \cite{bib:swgo1,bib:swgo2}, which will be realized at high altitude in the Southern Hemisphere, to be complementary to other gamma-ray experiments in the Northern Hemisphere, like HAWC \cite{bib:hawc} and LHAASO \cite{bib:lhaaso}, for the observation of the entire sky. It will operate with close to 100\% duty cycle and order steradian field of view. The site has to be at high altitude (above $4\,400$~m~a.s.l.), in order to be closer to the maximum of the extensive air showers induced by astrophysical gamma-rays with primary energy in the range of interest (between 100~GeV and a few PeV). The SWGO design will be primarily based on water Cherenkov detectors, and the final array and detector unit configurations are still to be defined \cite{bib:swgo1}. One configuration under study consists of and array of (surface) water Cherenkov tanks arranged in a high fill-factor core (with area considerably larger than HAWC) and a low density outer array. 

To study the single tank behaviour, we performed simulations of particles crossing tanks with different size and configuration of PMTs. We simulated double-layer tanks \cite{bib:dlt}, in which the lower layer helps in the gamma/hadron discrimination, as muons are more abundant in hadronic showers and they can cross the upper layer reaching the lower layer where they are measured. We considered tanks of different shape, with circular (Circular-DLT), hexagonal (Hexagonal-DLT) and square (Square-DLT) base. 

To simulate the particles crossing the tanks and their response we used the HAWCSim framework \cite{bib:hawcsim}, which makes use of GEANT4 \cite{bib:geant4} to simulate the interaction of the particle with the tank itself and the water volume, including the production of the Cherenkov photons that can be detected by the PMTs inside the tank.

\section{Simulations}

\subsection{Particles}

In this analysis we considered the most abundant particles contained in an extensive air shower generated by 400~GeV protons and 200~GeV photons at an observation level of $4\,500-5\,000$~m~a.s.l..  Therefore, we performed simulations of electrons, gamma-rays and muons with fixed energies: 10~MeV, 100~MeV and 1~GeV electrons and photons, and 1~GeV and 10~GeV muons. 
To define the directions, we used azimuth angles $\phi$ uniformly distributed in the range $0-360\deg$ and zenith angles $\theta$ in the range $0-60\deg$ sampled on a $\cos^{2}\theta$ distribution.

The particles were generated on a large circular surface 10~cm above the tank and centered with it. The size of the generation area is such that even the most inclined particles could enter the tank from the lateral walls of the upper layer, to avoid the detection of particles entering in the tank directly from the lower layer that would affect the overall performance of a single tank study. This has to be considered in the context of a sparse array of tanks, while in a dense array the nearby tanks contribute to the detection capability of the large scale experiment. Therefore, for each tank design, particle type and energy, 10000 particles entering the upper layer of the tanks have been analyzed.

\subsection{Specifications of the tanks}

\subsubsection{Shapes and dimensions of the tanks}

\begin{figure*}[h!]
	\centering
	\includegraphics[width=0.3\columnwidth]{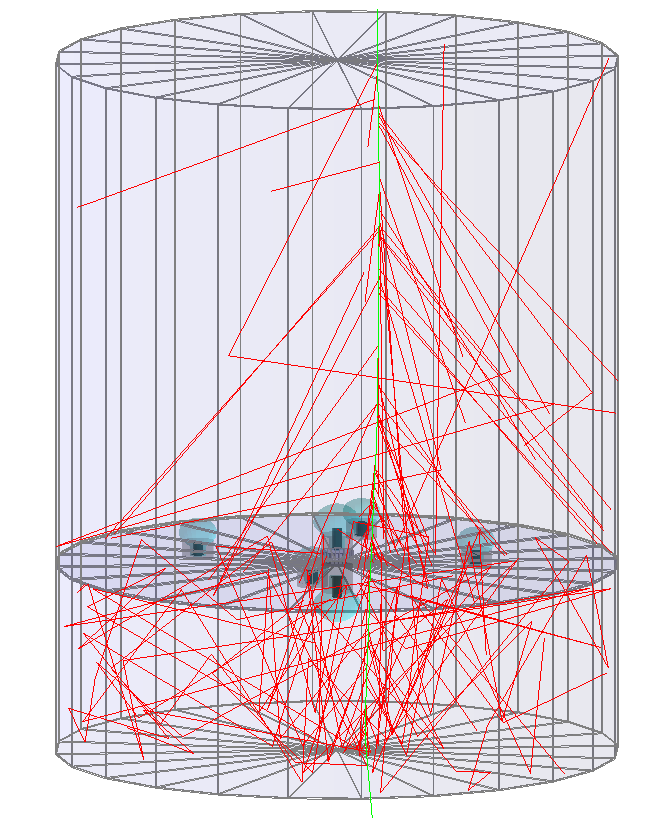}
	\includegraphics[width=0.28\columnwidth]{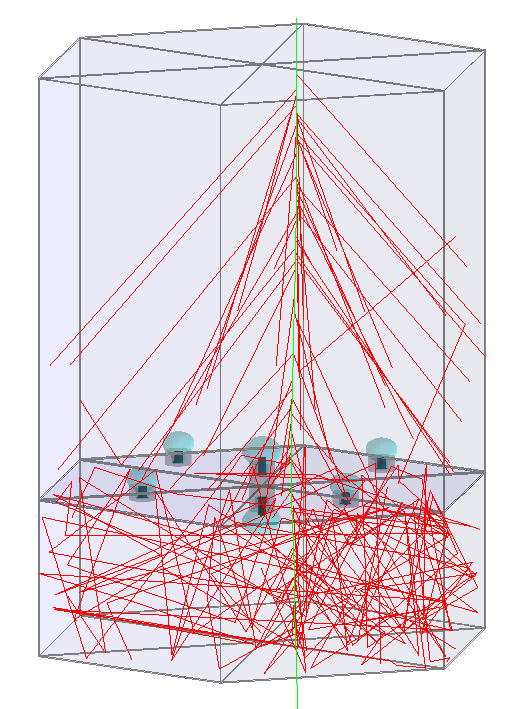}
	\includegraphics[width=0.32\columnwidth]{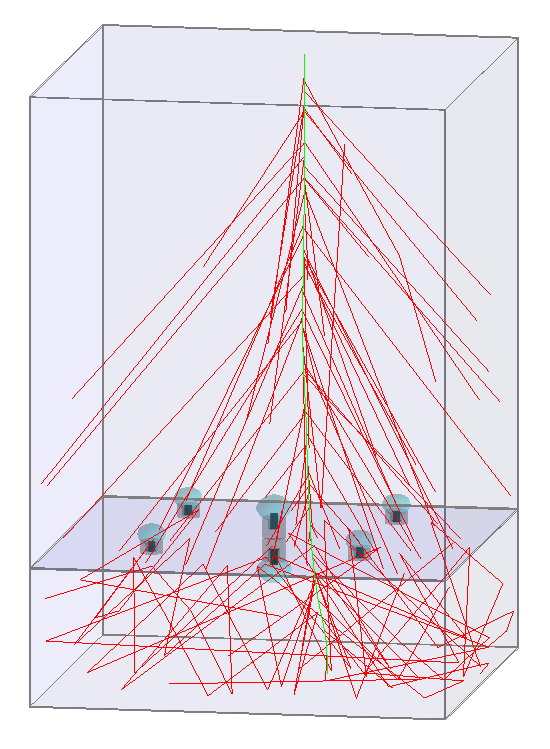}
	\caption{Geant4 visualization of a Circular-DLT, an Hexagonal-DLT and a Square-DLT, crossed by a 1~GeV vertical muon. All tanks have widths of 3~m (diameter for Circular-DLT, two  times  the  side  for  Hexagonal-DLT, and side for Square-DLT) and lower layers 1~m high. The upper layers were simulated with non-reflective walls, while the lower layer with reflective walls. The green line represents the simulated muon and the red lines a sample of Cherenkov photons.} \label{fig:tanks_g4}
\end{figure*}

In this analysis, Circular-DLTs, Hexagonal-DLTs and Square-DLTs were considered. In Fig.~\ref{fig:tanks_g4} examples of the Geant4 visualization of the three tank designs crossed by a muon are shown. 

The height of the upper layer was chosen allowing the Cherenkov photons to reach any PMT at the base of the upper layer. Assuming a vertical particle entering the tank from the center of the roof, the Cherenkov photons should be able to reach the lateral walls of a Circular-DLT or the corners of an Hexagonal-DLT and a Square-DLT at the base of the upper layer. 
For Circular-DLTs the height $h$ and radius $r$ follow the relation $h=r/\tan{\theta_C}$, where $\theta_C=41.2\deg$ is the emission angle of the Cherenkov photons with respect to the trajectory of the particle crossing the water. Similarly, for Hexagonal-DLTs with side $L$, the height is $h=L/\tan{\theta_C}$, and for Square-DLTs with half side $l$, $h=\sqrt{2}l/\tan{\theta_C}$. To the height calculated with previous formulas, 1~m of water is added to have 90\% probability that gamma-rays interact by pair production. The lower layer, with height independent of the radius, is dedicated to muon measurements, allowing for the gamma/hadron discrimination and the separation of mass groups of charged primaries (from 2 to 4). For the lower layer, we chose heights of 0.5~m, 0.75~m and 1~m. The dimensions of the tanks are collected in Tab.~\ref{tab:size_tanks}.
\begin{table}[ht]
	\centering
	\small
	\begin{tabular}{|c|c|c|c|c|}
		Tank & Width (m) & Cir.\&Hex. Height u.l. (m) & Sqr. Height u.l. (m) & Height l.l. (m) \\
		\hline
		T1  & 3    &  2.7  &  3.4  &  0.5, 0.75, 1 \\
		T2  & 3.5  &  3.0  &  3.8  &  0.5, 0.75, 1 \\
		T3  & 4    &  3.3  &  4.2  &  0.5, 0.75, 1 \\
		T4  & 4.5  &  3.6  &  4.6  &  0.5, 0.75, 1 \\
		T5  & 5    &  3.9  &  5.0  &  0.5, 0.75, 1 \\
		T6  & 5.5  &  4.2  &  5.4  &  0.5, 0.75, 1 \\
		
	\end{tabular}
	\caption{Size of the tanks. ``Width'' is the diameter of Circular-DLT, the side of Square-DLT and two times the side of Hexagonal-DLT; ``Cyl.\&Hex.~Height u.l.'' is the height of the upper layer of Circular-DLT and Hexagonal-DLT; ``Sqr.~Height u.l.'' is the height of the upper layer of Square-DLT; ``Height~l.l.'' is the height of the lower layer.}\label{tab:size_tanks}
\end{table}

\subsubsection{Properties of the inner walls}

For the inner walls of the upper layers, we used both reflective (Tyvek) and non-reflective (Polypropylene) materials. The reflectivity of the materials depends on the wavelength of the incident photons. Tyvek has a reflectivity of $0.63-0.92$ in the wavelength range $250-650$~nm; polypropylene has a reflectivity of 0.10 over the same wavelength range. Reflective walls allow for a better detection capability, but might extend the detection time, due to possible consecutive reflections of photons on the walls before they reach the PMTs. This results in a higher detection efficiency as photons that would not be detected with non-reflective walls are instead detected with reflective walls, but  also widen the time resolution for the detection of the first photon. For the lower layer we used reflective walls, as the priority was given to the detection efficiency of particles entering the lower layer rather than the timing.

\subsubsection{PMTs}

In the upper layer we used two configurations of PMTs looking upwards: one central 10\textquotedbl~PMT or four peripheral 5\textquotedbl~PMTs placed at half radius in Circular-DLTs, half the apothem in Hexagonal-DLTs, and half diagonal in the Square-DLT. Signals of the peripheral PMTs are summed in one unique output. In the lower layer we used one central 10\textquotedbl~PMT or 5\textquotedbl~PMT looking downwards. In each layer, the two PMT configurations have to be considered independently. In the simulations, we used two models of PMTs from Hamamatsu: the 10\textquotedbl~R7081HQE PMT, and the 8\textquotedbl~R5912 PMT, then re-scaled to a 5\textquotedbl~PMT during the analysis phase. 

\section{Analysis}\label{sec:analysis}

For the evaluation of the tank response, the parameters taken into account are:
\begin{figure}[th!]
	\centering
	\subfloat[e$^-$ - number of PEs]{\includegraphics[width=0.33\linewidth]{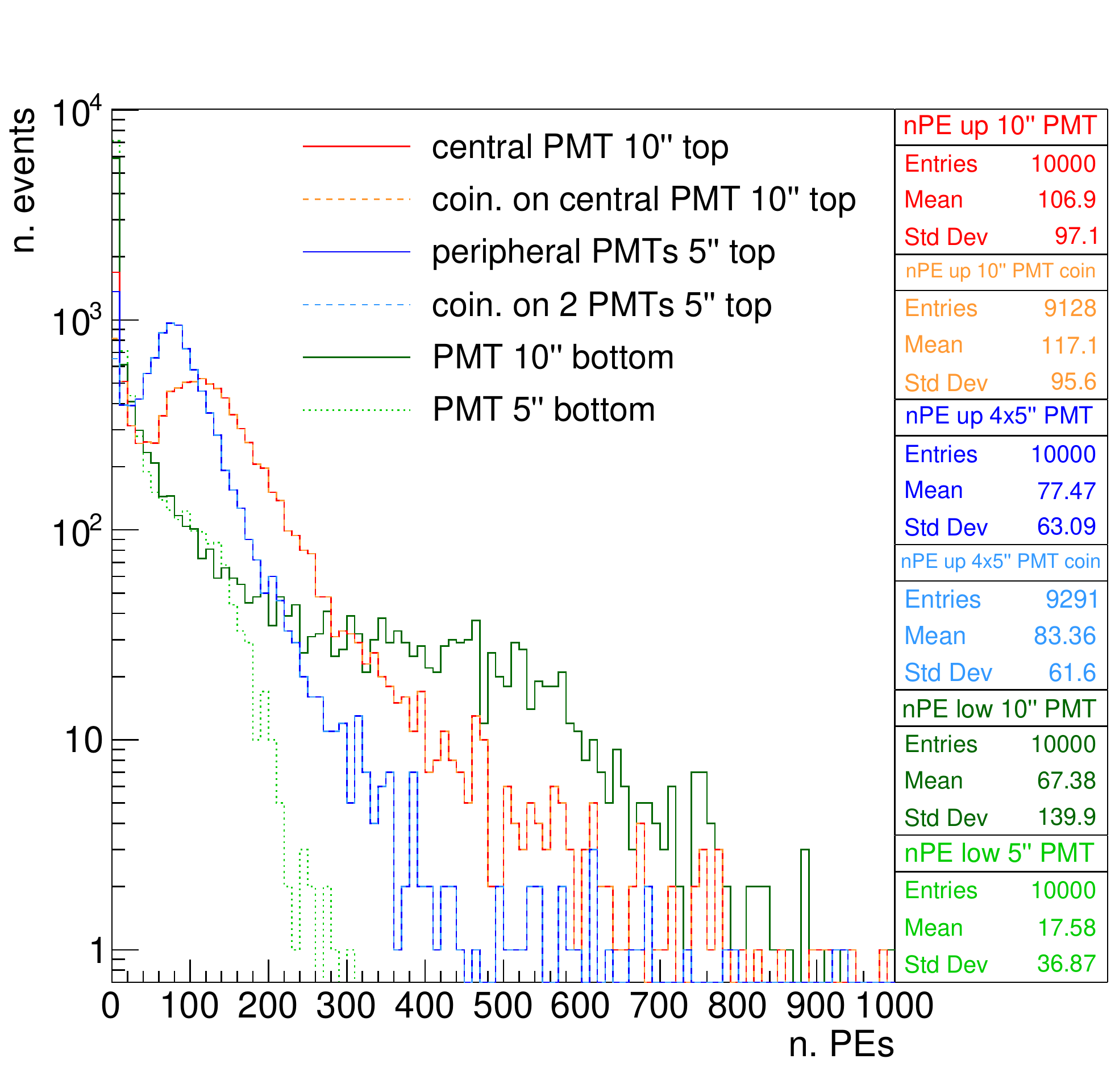}}
	\subfloat[$\gamma$ - number of PEs]{\includegraphics[width=0.33\linewidth]{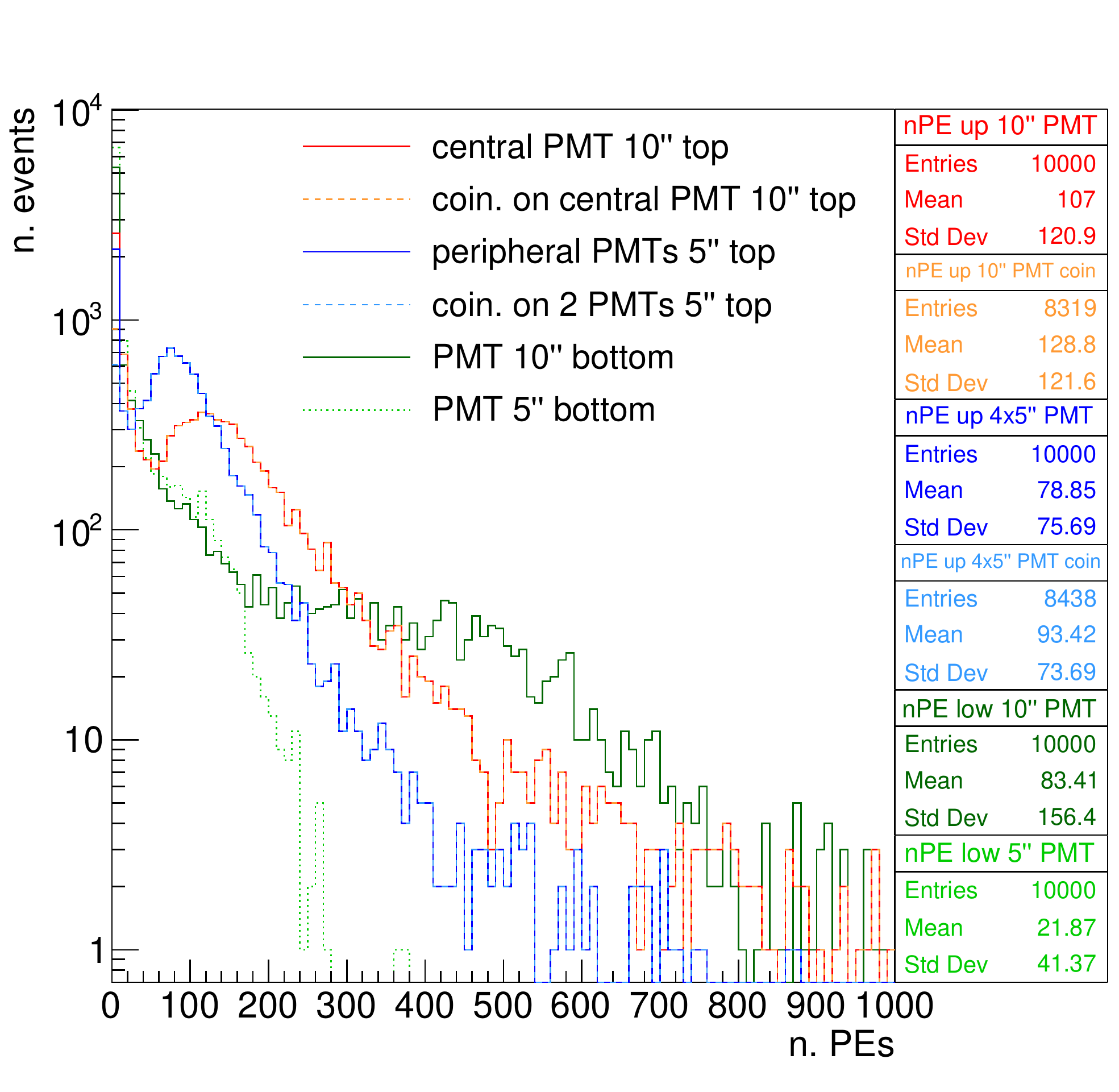}}
	\subfloat[$\mu^-$ - number of PEs]{\includegraphics[width=0.33\linewidth]{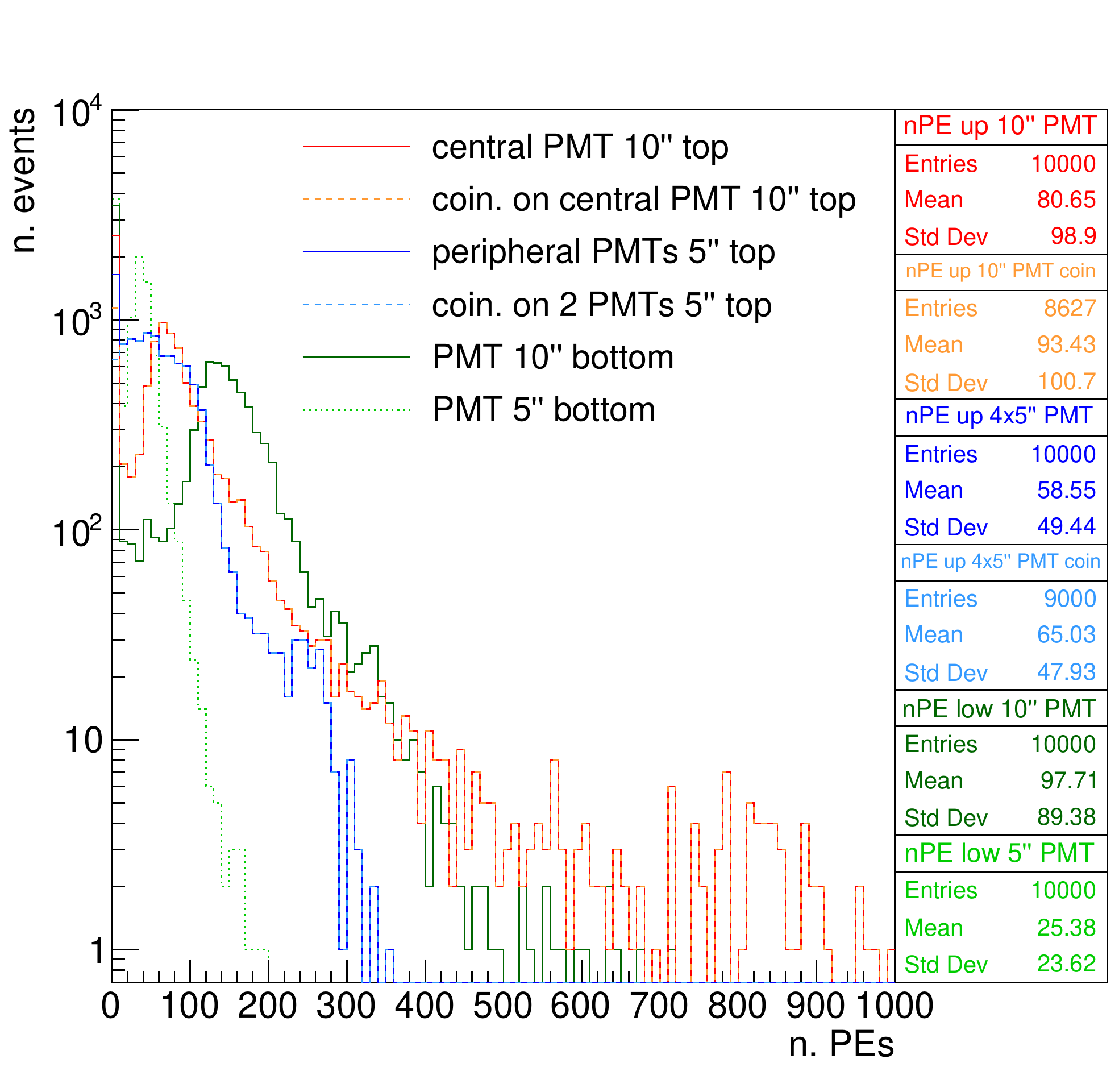}}\\
	\subfloat[e$^-$ - time first photon]{\includegraphics[width=0.33\linewidth]{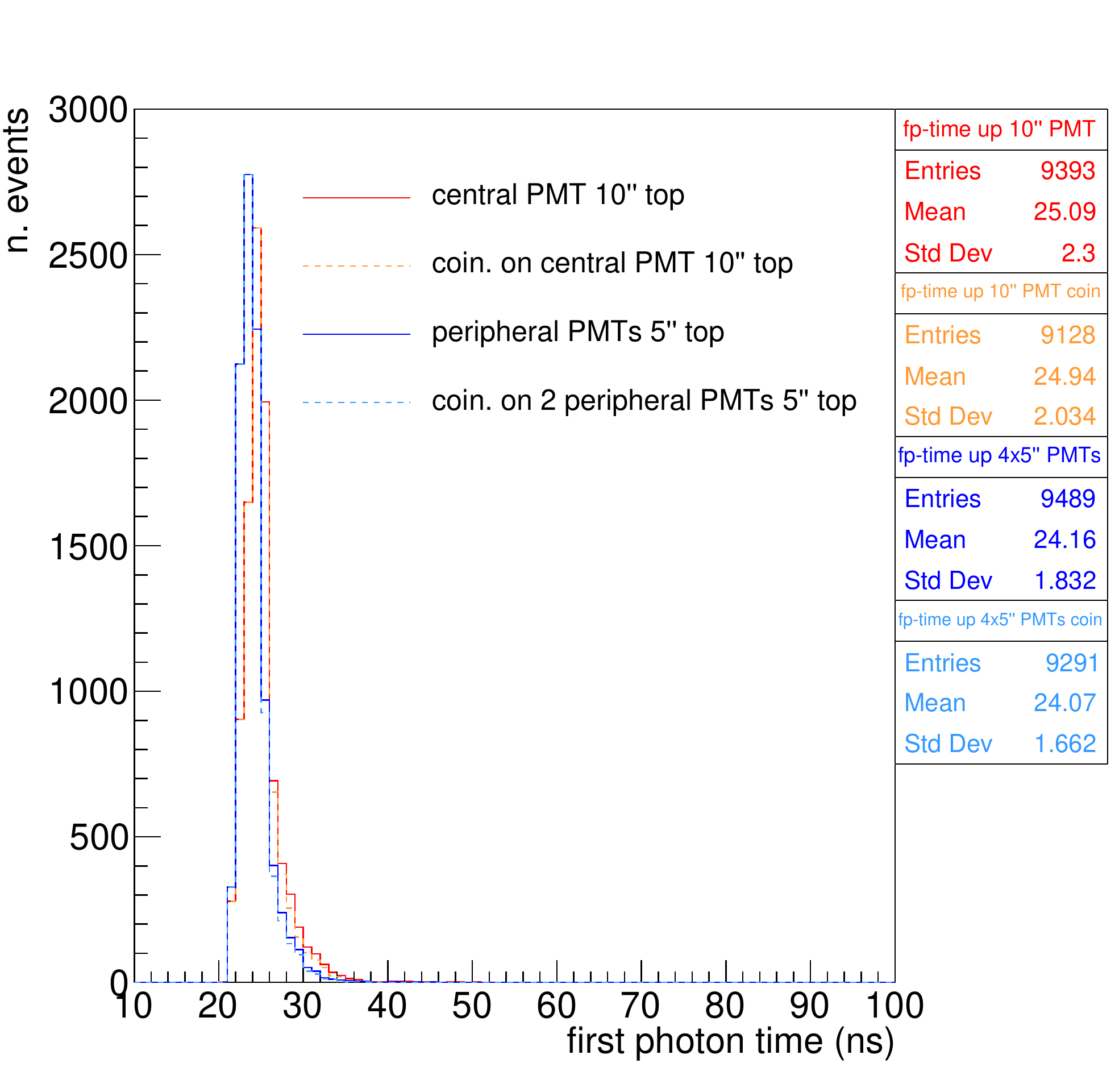}}
	\subfloat[$\gamma$ - time first photon]{\includegraphics[width=0.33\linewidth]{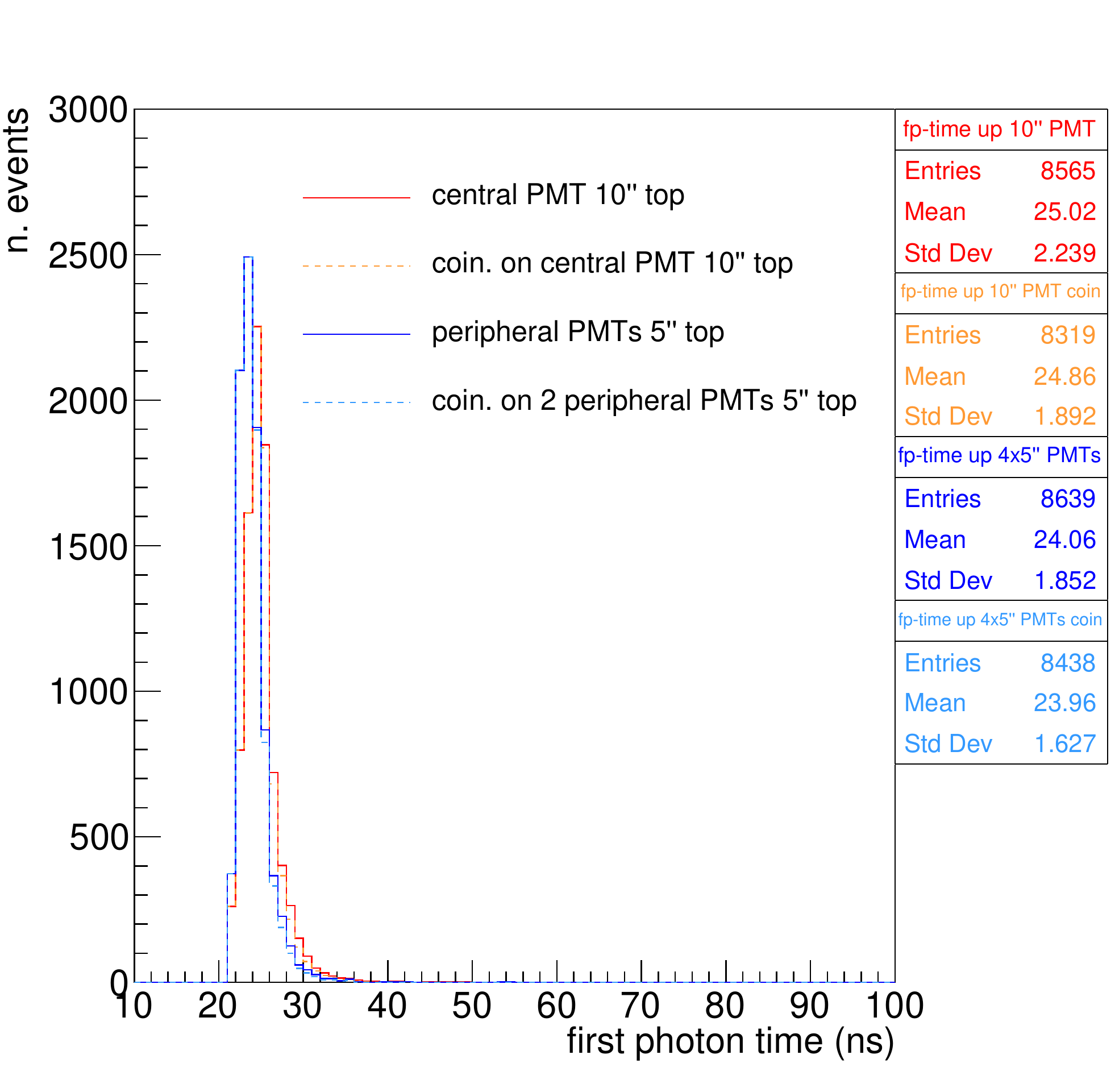}}
	\subfloat[$\mu^-$ - time first photon]{\includegraphics[width=0.33\linewidth]{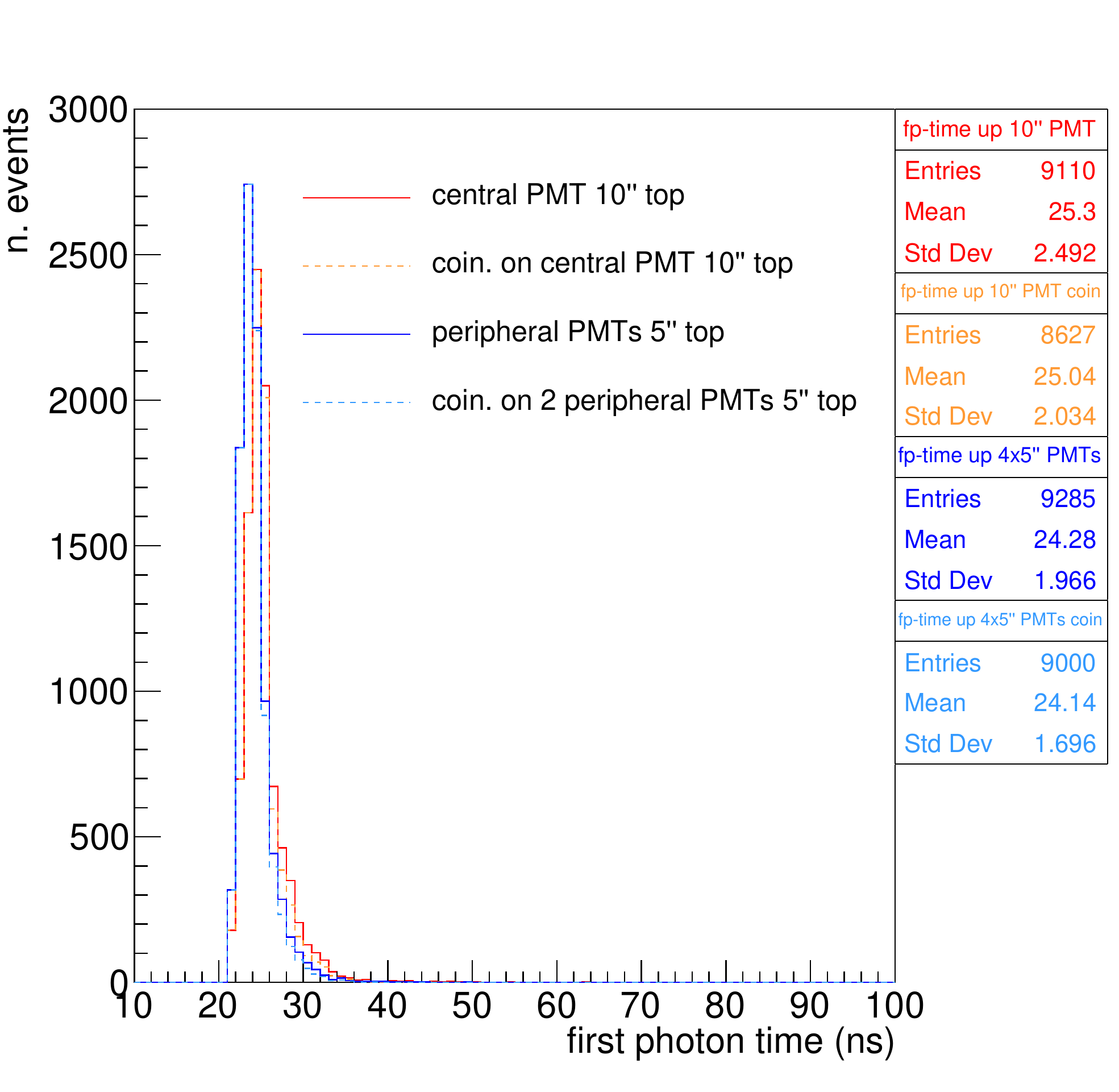}}\\
	\subfloat[e$^-$ - time]{\includegraphics[width=0.33\linewidth]{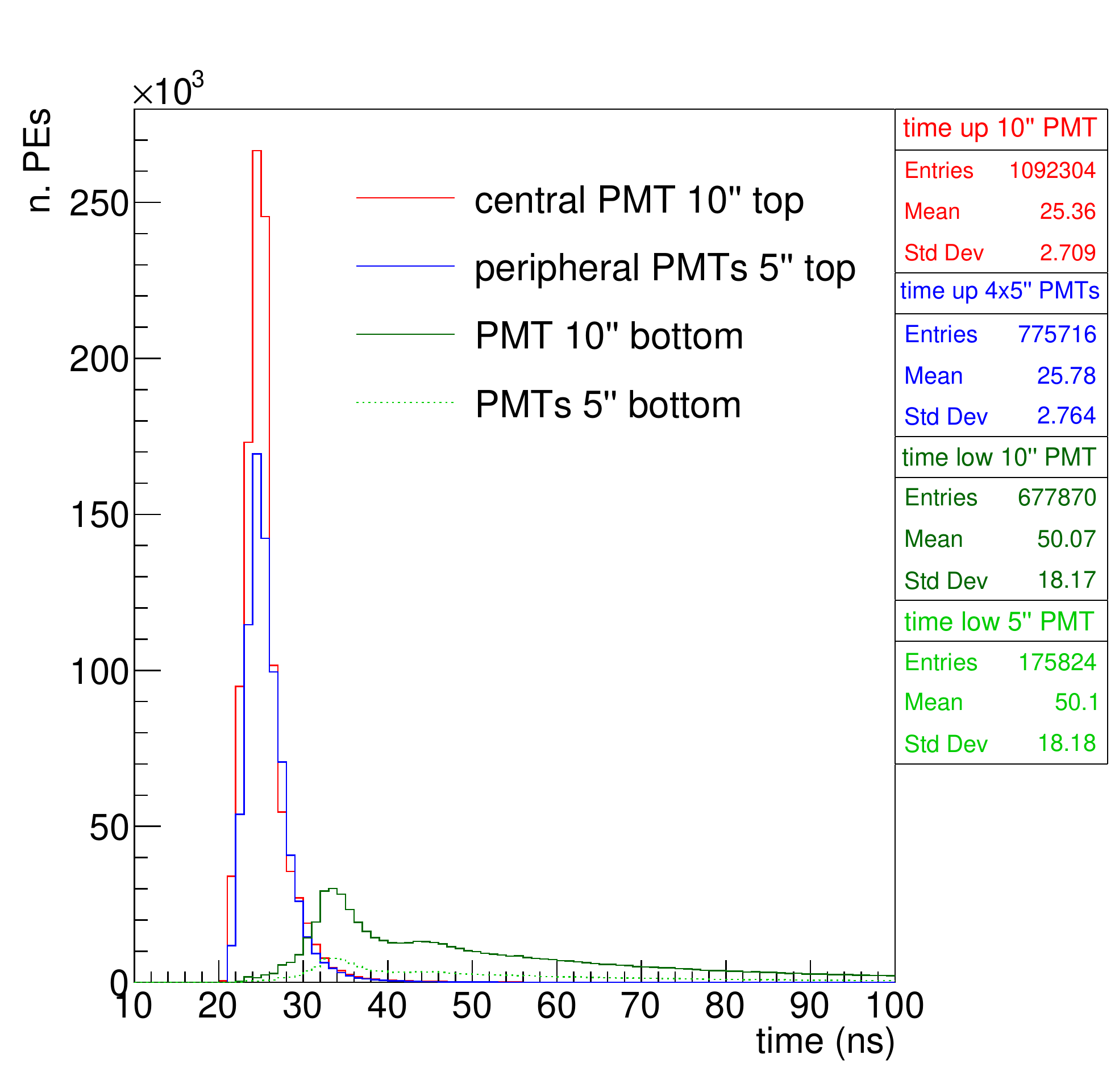}}
	\subfloat[$\gamma$ - time]{\includegraphics[width=0.33\linewidth]{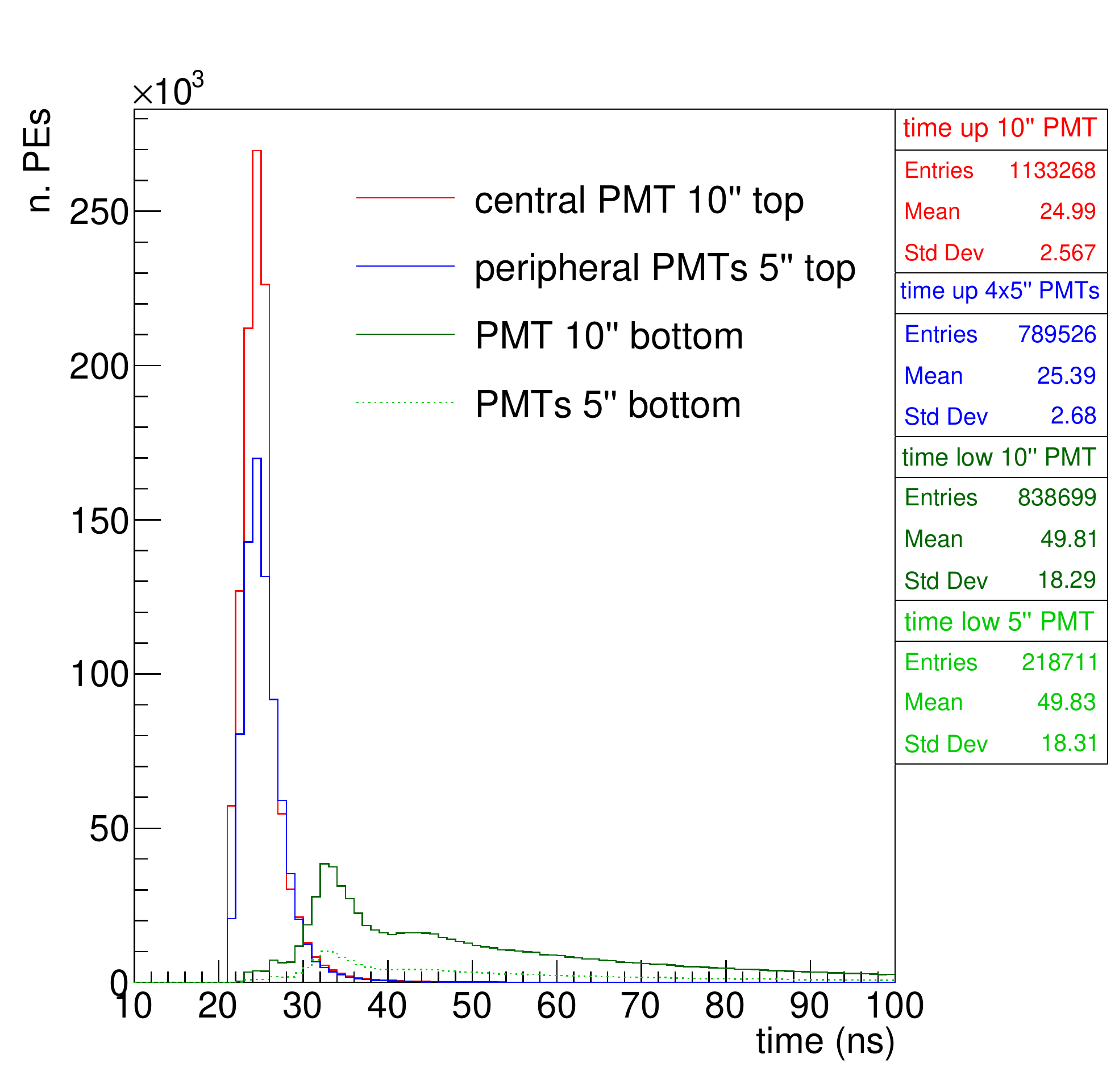}}
	\subfloat[$\mu^-$ - time]{\includegraphics[width=0.33\linewidth]{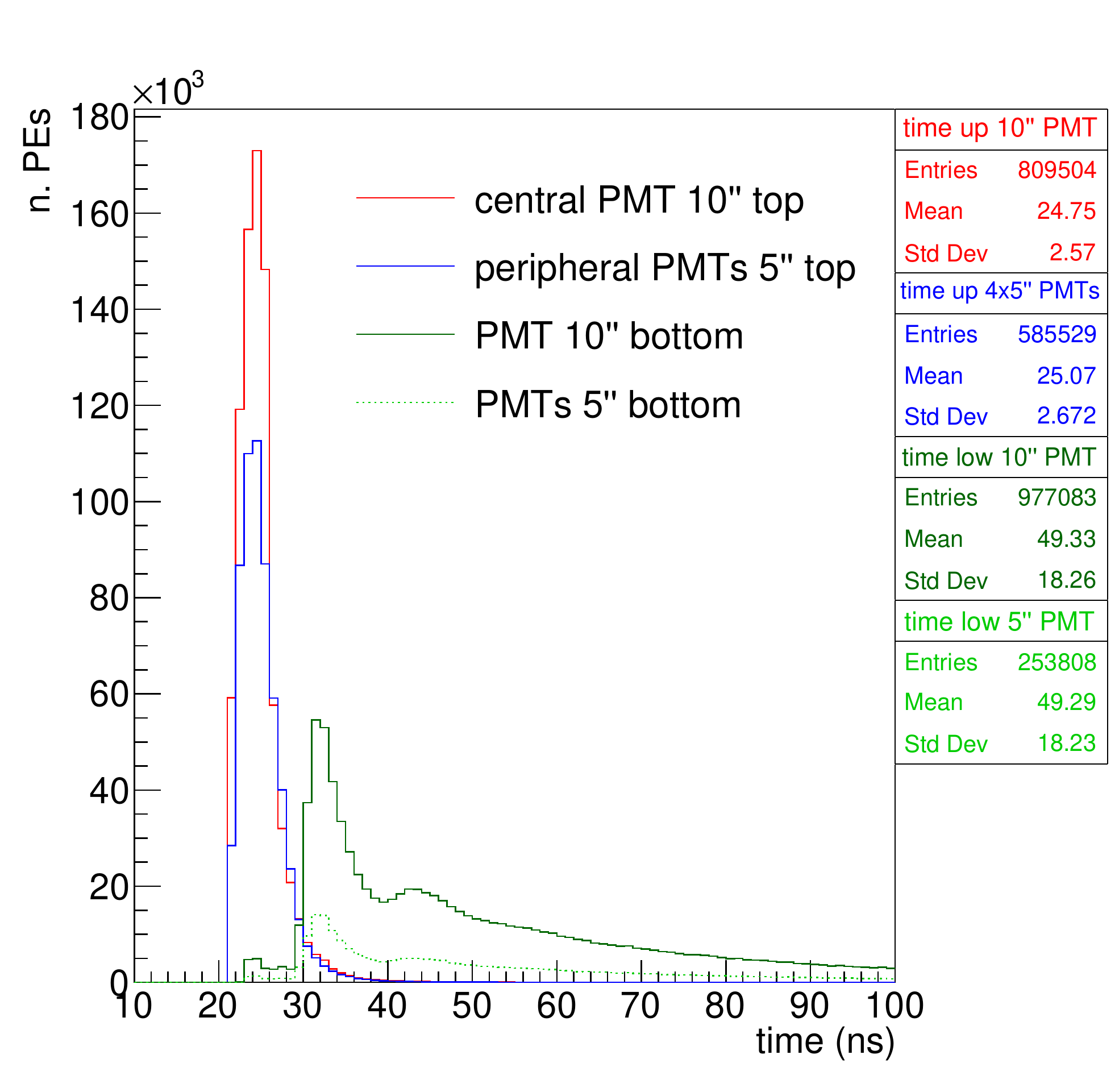}}
	
	\caption{Distributions used in the analysis of the tank performance: number of PEs (a-c), arrival time of the first photon (d-f), and arrival time of any photon (g-i). These plots refer to simulations of 1~GeV electrons, gamma-rays and muons crossing a Circular-DLT with non-reflective walls in the upper layer and reflective walls in the lower layer. The statistics boxes shown on the right-side of the plots can be used to analyze the results for the different configurations of PMTs. For example, the ``Mean'' value in (a-c) can be used to estimate the average number of PEs, while the width of the distribution ``Std~Dev'' in (d-f) can be used to evaluate the time resolution of the measurement of the first photon.} \label{fig:plots}
\end{figure}
\begin{itemize}
	\item The number of photoelectrons (PEs) produced in both layers. In the upper layer we considered separately the configuration with one central 10\textquotedbl~PMT or four peripheral 5\textquotedbl~PMTs. For the lower layer we considered individually a central 10\textquotedbl~PMT or 5\textquotedbl~PMT. This allowed to understand which of the two configurations in the upper layer gives the higher detection efficiency and the better time resolution of the first detected photon and, for the lower layer, how a different size of the PMT changes the performance.
	\item The time resolution of the measurement of the first photon in the upper layer, evaluated as the standard deviation of the distribution of the first photon arrival time.
	\item The detection efficiency of both layers. The efficiency is calculated as the number of detected particles (events) divided by the number of particles entering the upper layer of the tank (10000). The latter is based on simple geometrical considerations, based on the height of the entrance point of the particles. Due to the random direction of particles, a fraction of the particles that enter the upper layer does not enter the lower layer. We tried to evaluate the number of particles actually entering the lower layer based on the initial direction of the particles, but this was not possible due to non-tracked deflections of the particle trajectories occurring while they cross the tank. 
	We performed a set of simulations where only 10~GeV vertical muons were thrown through a Circular-DLT, and in this case the detection efficiency was $\sim$1 for both the upper and lower layer. This demonstrated that the inefficiency of the lower layer is only due to geometrical constraints, and would be effectively reduced considering a joint detection of inclined particles by neighbouring tanks in a dense array. Therefore, also in the calculation of the detection efficiency of the lower level we used as reference the number of particles entering the upper layer of the tank. This effect underestimates the detection efficiency of the lower layer for any type of tank and particle, but the comparison between the different configurations remains valid. For the upper layer, we considered as threshold both 1~PE and the coincidence of 2~PEs produced within 30~ns by the central 10\textquotedbl~PMT or by the four peripheral 5\textquotedbl~PMTs, while for the lower layer the threshold was only of 1~PE.
\end{itemize}

In Fig.~\ref{fig:plots} sample distributions of the number of PEs (a-c), the arrival time of the first photon (d-f), and the arrival time of photons (g-i) of  are shown. They refer to simulations of 1~GeV electrons, gamma-rays and muons crossing a Circular-DLT with non-reflective walls in the upper layer and reflective walls in the lower layer. 
The distribution of the number of PEs in the lower layer has a higher average for muons than for electrons and gamma-rays, as the latter are absorbed by the water of the upper layer. The PE distributions also show that particles generate in average more PEs on the central 10\textquotedbl~PMT than on the four peripheral 5\textquotedbl~PMTs. 
The distributions of the arrival time of the first photon are similar. In the sample of distributions reported, the timing resolution for the four peripheral PMTs is slightly narrower than for the central PMT.

The time distributions of photons are also presented to show the effect of the reflective walls in the lower layer. Especially for muons that reach easily the lower layer, a long tail and a bump after the main peak are due to the consecutive reflections on the walls before photons reach the PMT. This effect is more visible when reflective covers are used also for the upper layer, due to the higher statistics of PEs generated in the upper layer. Depending on the distributions of the impact point on the tank and the direction of the particles, a sequence of bumps might appear instead of the tail.

\section{Results}\label{sec:results}

\subsection{Comparison of tanks with different size, reflective properties of the inner walls and PMT configuration}

\begin{figure*}[h!]
	\centering
	\subfloat[]{\includegraphics[width=0.49\linewidth]{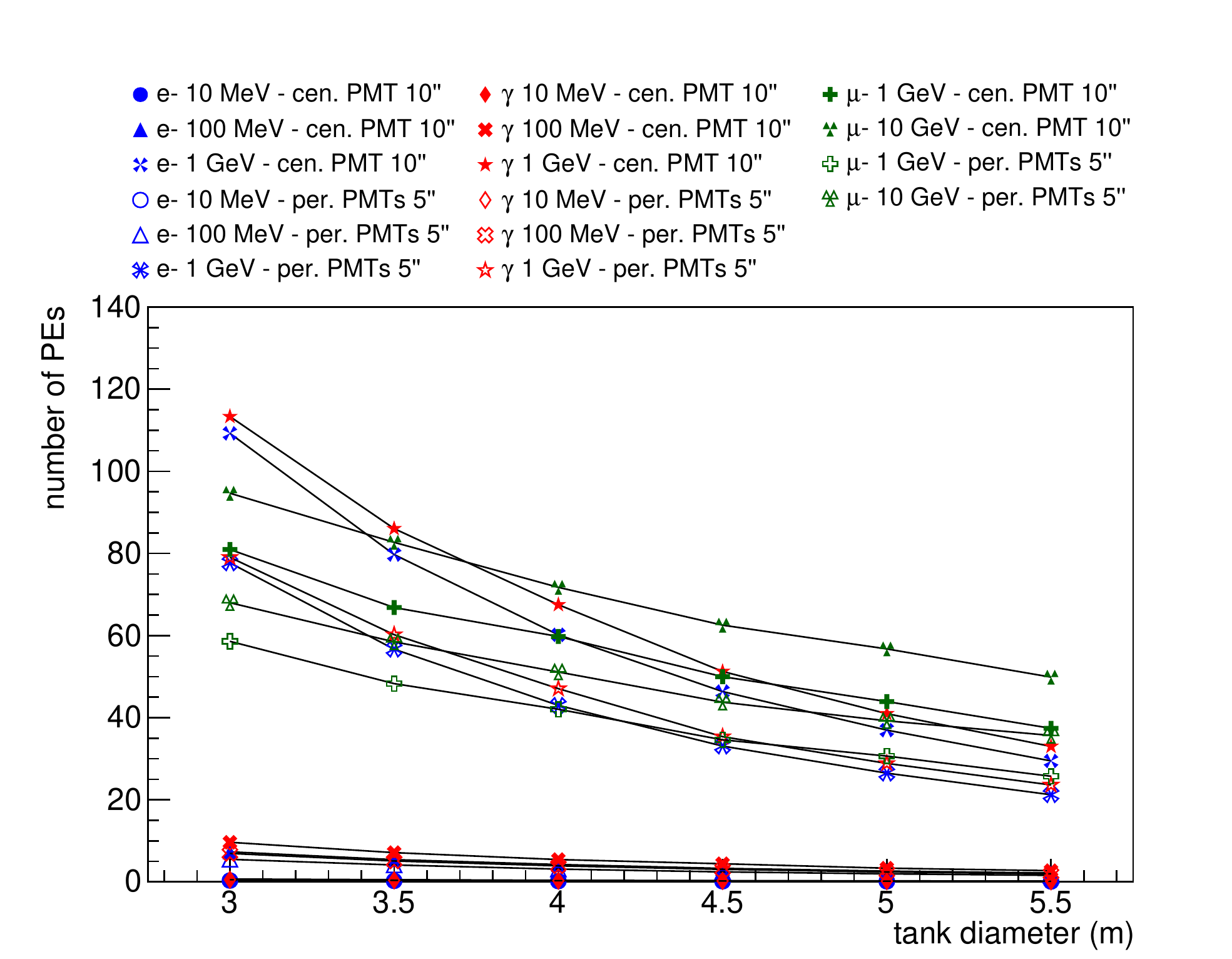}}
	\subfloat[]{\includegraphics[width=0.49\linewidth]{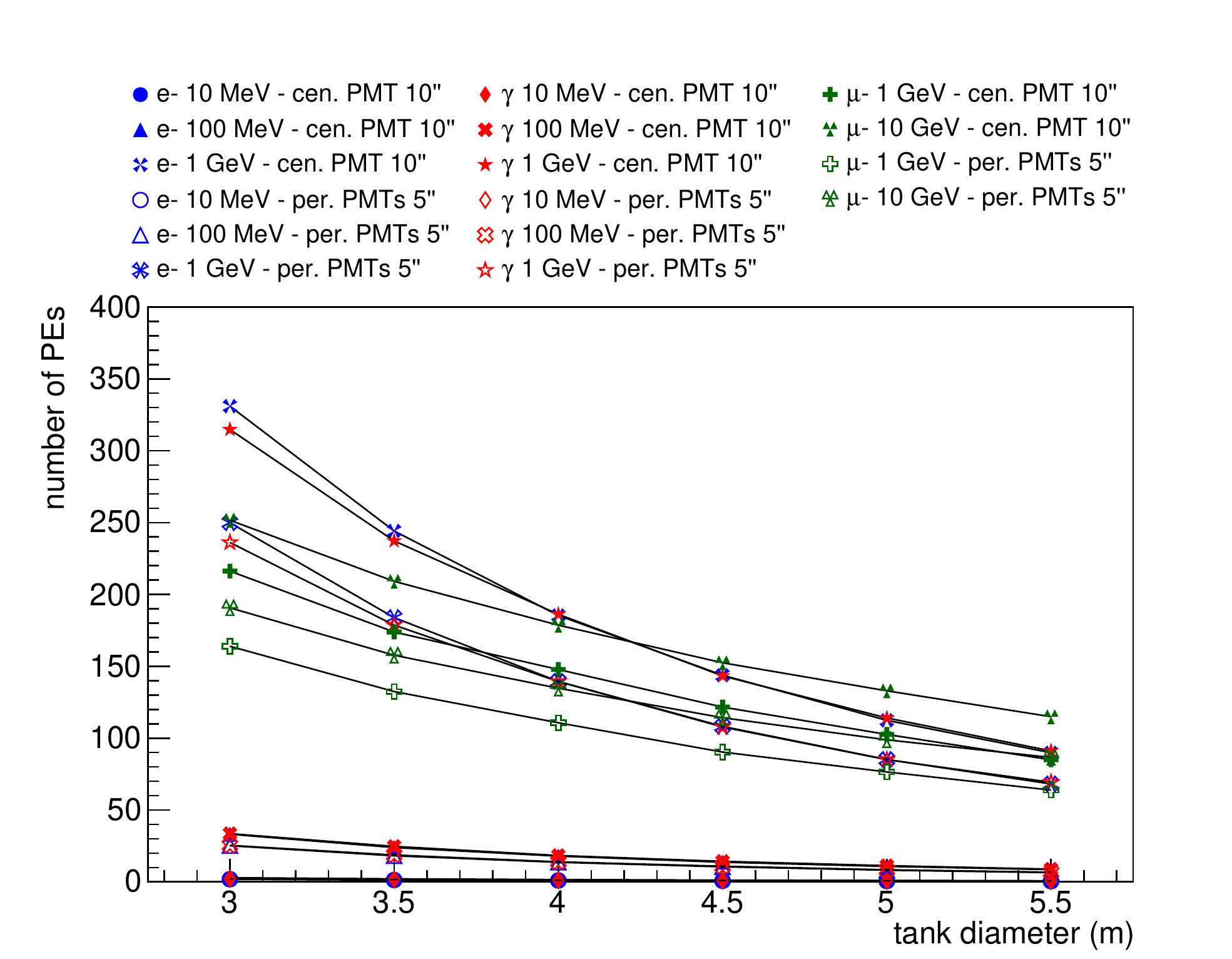}}
	\caption{Number of PEs detected in the upper layer with non-reflecting walls (a) and reflective walls (b) in Circular-DLT, note the different scales on the y-axis.} \label{fig:npe_up}
\end{figure*}

\begin{figure*}[h!]
	\centering
	\subfloat[]{\includegraphics[width=0.49\linewidth]{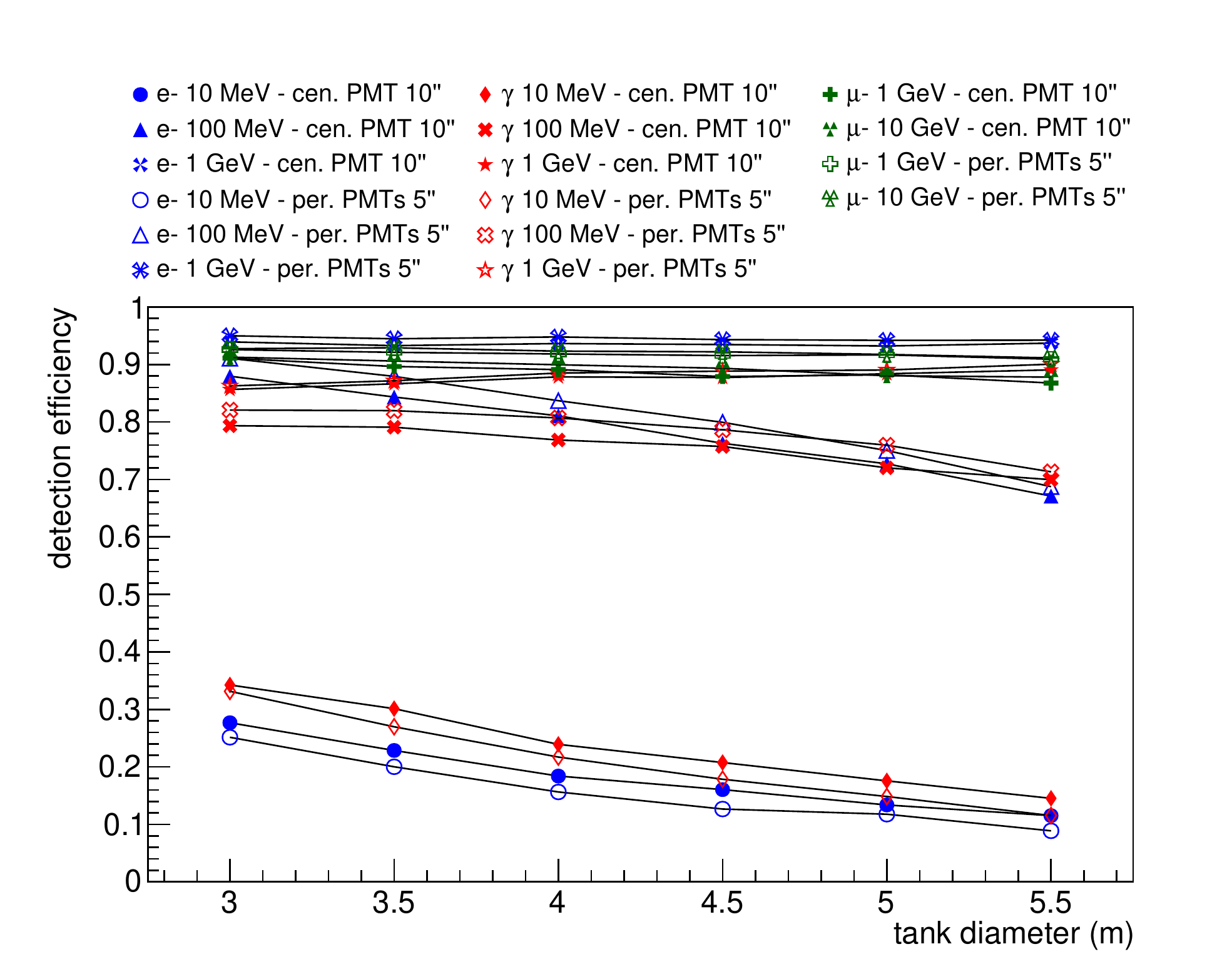}}
	\subfloat[]{\includegraphics[width=0.49\linewidth]{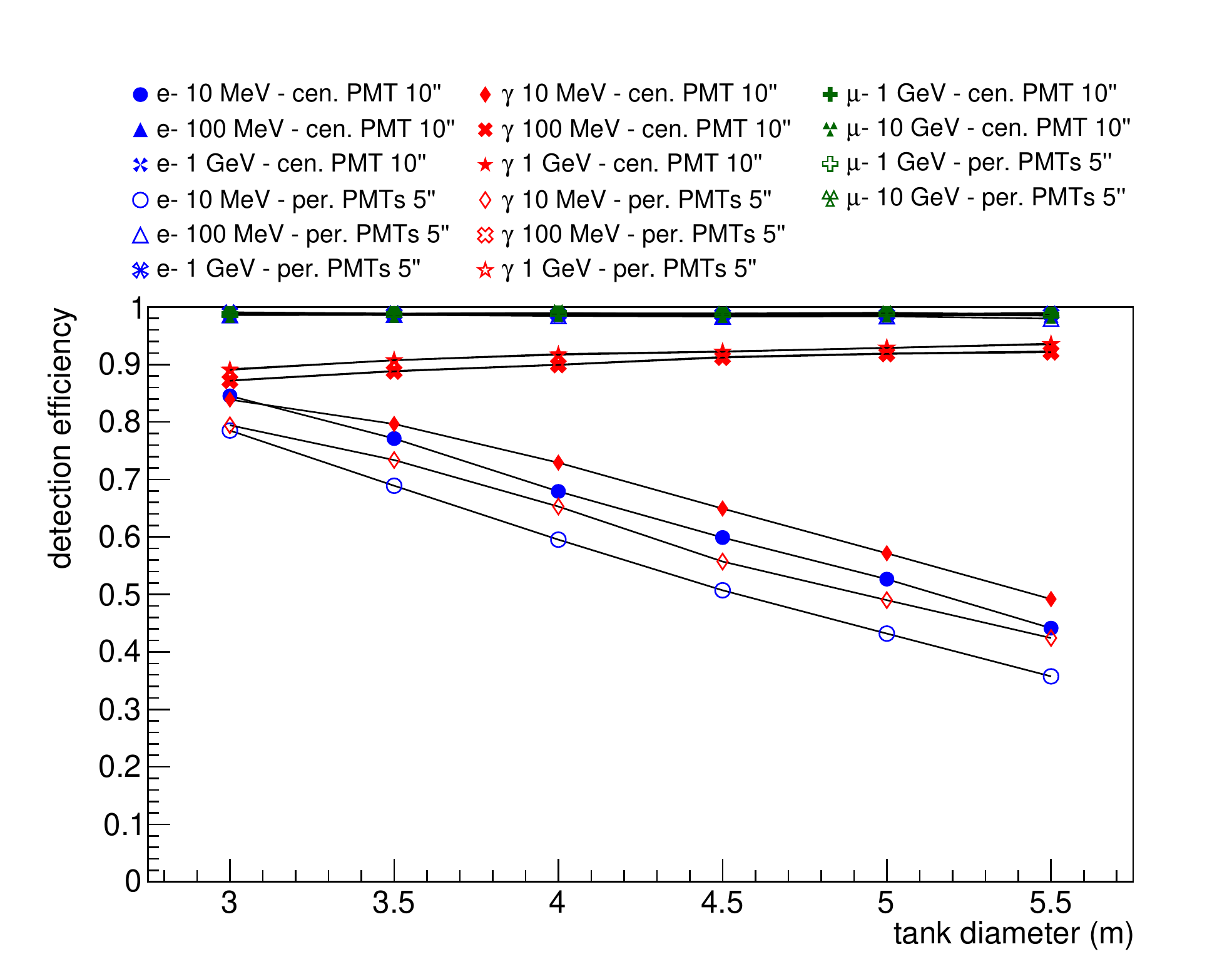}}
	\caption{Detection efficiency of the upper layer with non-reflecting walls (a) and reflective walls (b) in Circular-DLT.} \label{fig:eff_up}
\end{figure*}

\begin{figure*}[h!]
	\centering
	\subfloat[]{\includegraphics[width=0.49\linewidth]{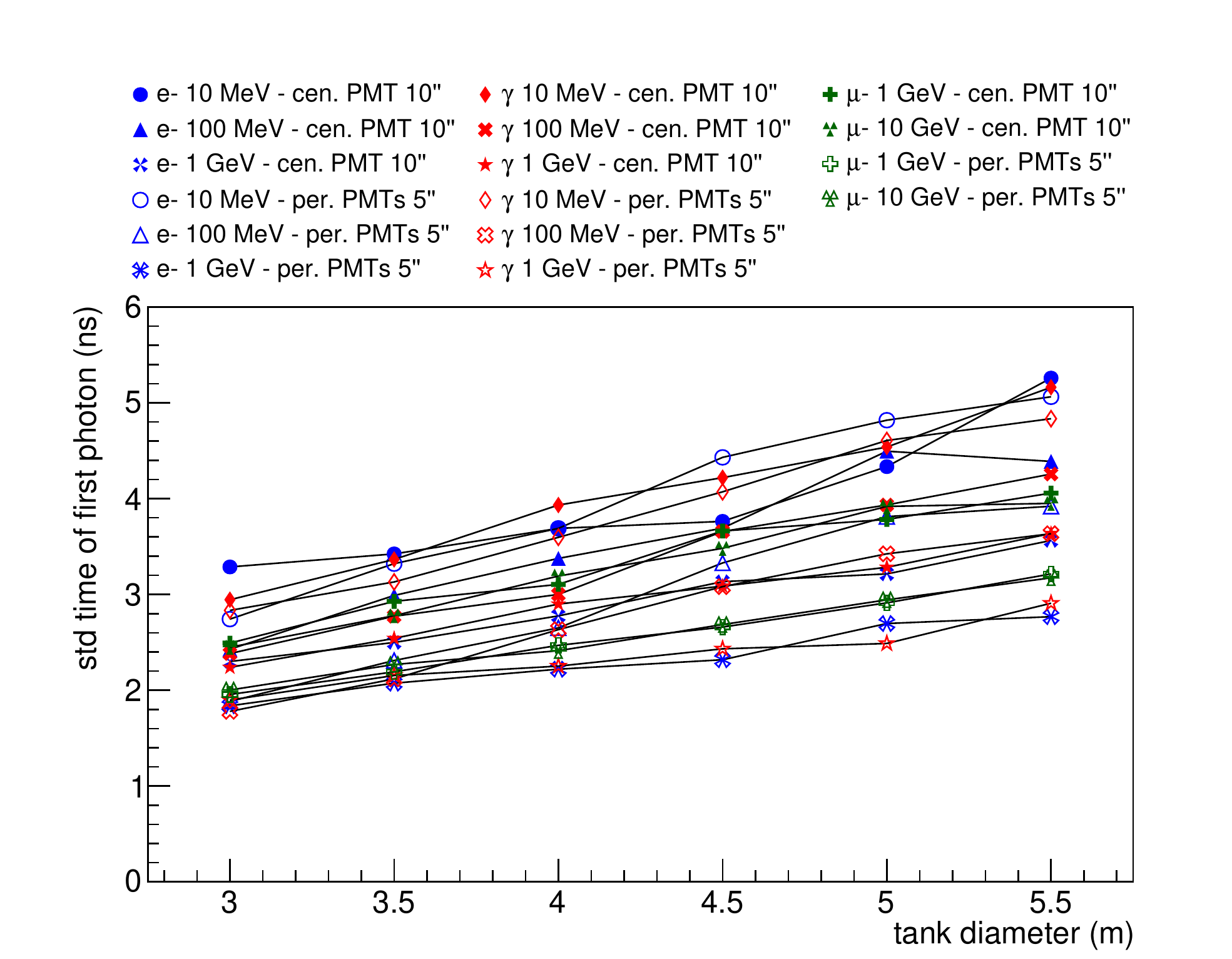}}
	\subfloat[]{\includegraphics[width=0.49\linewidth]{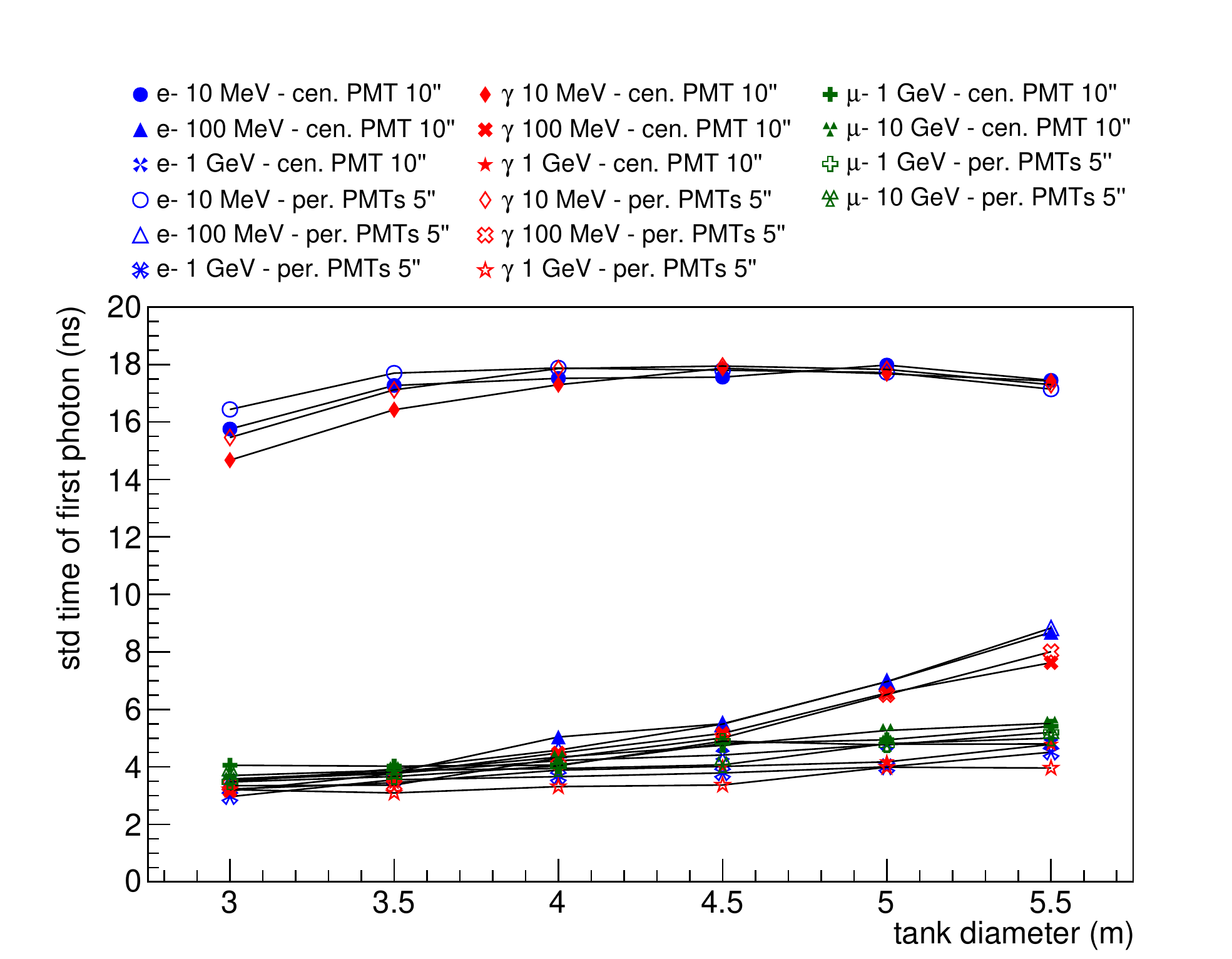}}
	\caption{Time resolution of the measurement of the first photon in the upper layer with non-reflective walls (a) and reflecting walls (b) in Circular-DLT.} \label{fig:std_ftime}
\end{figure*}

\begin{figure*}[ht]
	\centering
	\subfloat[]{\includegraphics[width=0.49\linewidth]{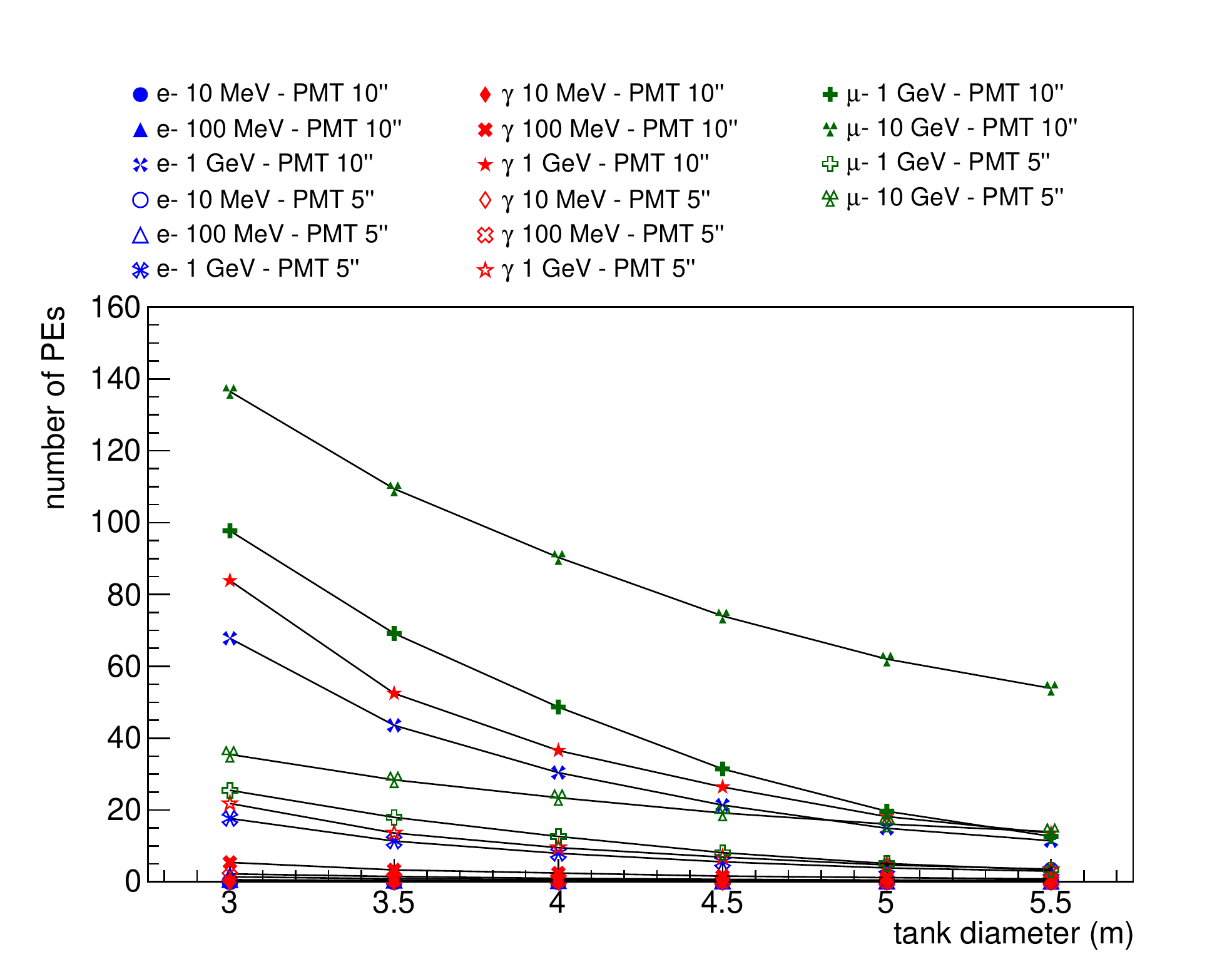}}
	\caption{Number of PEs detected in the lower layer with reflective walls.} \label{fig:npe_low}
\end{figure*}

\begin{figure*}[h!]
	\centering
	\subfloat[]{\includegraphics[width=0.49\linewidth]{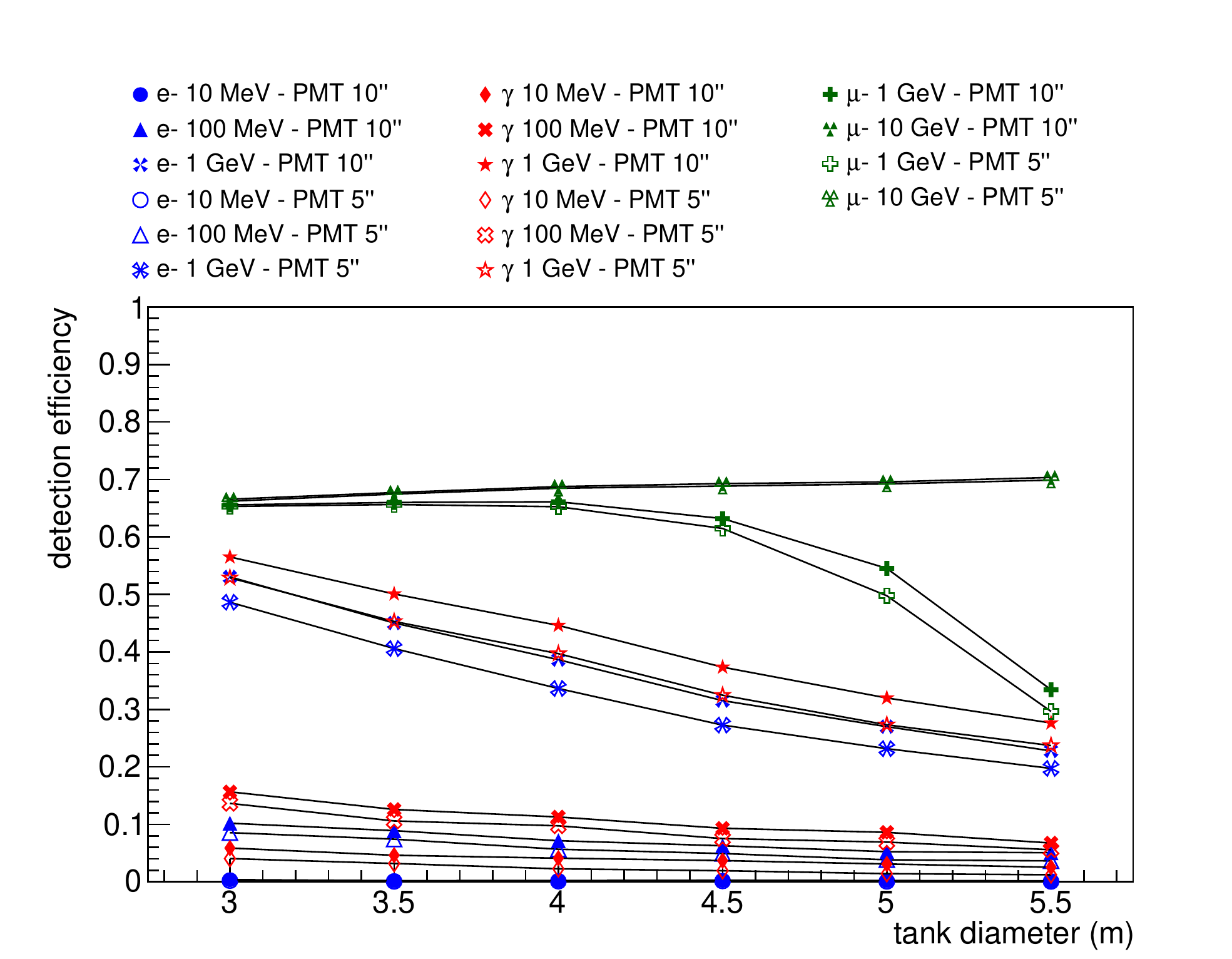}}
	\caption{Detection efficiency of the lower layer with reflective walls. In all the configurations, it is underestimated due to geometrical constraints.} \label{fig:eff_low}
\end{figure*}

In this section, plots are referred to Circular-DLTs, but the results are similar for all tank geometries. In Fig.~\ref{fig:npe_up}--\ref{fig:std_ftime} the performance of the upper layer is shown, considering a central 10\textquotedbl~PMT or four peripheral 5\textquotedbl~PMTs. Panels (a) are relative to non-reflecting walls, while panels (b) are for reflecting walls. Similarly, in Fig.~\ref{fig:npe_low}--\ref{fig:eff_low} the performance of the lower layer, which has always reflective walls, is shown considering a 10\textquotedbl~PMT or a 5\textquotedbl~PMT. 

The number of detected PEs in the upper layers (see Fig.~\ref{fig:npe_up}) and consequently the detection efficiency (see Fig.~\ref{fig:eff_up}) decrease while increasing the size of the tank. This is due to the decrease of the ratio between the area of the PMT and that of the base of the tank. To verify this, we made some test simulations using different tank widths and rescaling the PMT size in order to have a constant ratio between the area of the PMT and the base of the tank, and the detection efficiency remained almost constant.
With 1~PE threshold, the detection efficiency of the upper layers considering one central 10\textquotedbl~PMT or four peripheral 5\textquotedbl~PMTs are comparable, although more PEs are produced in the central PMT. The sensitive area based on the size of the PMTs is similar for the two configurations. Therefore, the difference is related to the position of the PMTs. The efficiency for 10~MeV and 100~MeV particles is a few ten percent, while it is above 70\% for higher energies. With 2~PEs threshold (plots not shown in this work), the efficiency is reduced by a few ten percent for 10~MeV and 100~MeV particles, while it is similar for particles with higher energy. 
With reflective walls in the upper layers, the number of PEs and the detection efficiency are higher than those for non-reflective walls, even for low energy particles (compare  Fig.~\ref{fig:npe_up} (a) with (b) and Fig.~\ref{fig:eff_up} (a) with (b)). Using 2~PEs threshold, the effect is the same as that for non-reflecting walls.

The time resolution of the measurement of the first photon worsen, i.e. the standard deviation of the distribution gets larger, with the increase of the size of the tank (see Fig.~\ref{fig:std_ftime}). In average, using non-reflecting walls it ranges from $\sim$2.5~ns in small tanks to $\sim$3.5~ns in large tanks.
Considering 2~PEs threshold, it has smaller values, between $\sim$1~ns and $\sim$2~ns.
Using reflective walls, it slightly increases for 100~MeV and 1~GeV particles, and rises up to $\sim$18~ns for particles of 10~MeV, because the time distribution of the first photon shows a long tail for these particles. It has similar values for the central 10\textquotedbl~PMT and the four peripheral 5\textquotedbl~PMTs. Considering 2~PEs threshold, it has slightly lower values.

Like in the upper layer, the number of detected PEs (see Fig.~\ref{fig:npe_low}) and the detection efficiency (see Fig.~\ref{fig:eff_low}) in the lower layers decrease with the size of the tank. Electrons and gamma-rays of 10~MeV and 100~MeV are rarely detected in lower layers. Considering a 5\textquotedbl~PMT instead of a 10\textquotedbl~PMT, the efficiency slightly decreases although the number of produced PEs is the 25\%, proportional to the area of the photocathode. Both kinds of PMTs are placed at the center of the ceiling of the lower layer, so there is no effect due to different positioning like it happens in the upper layer.
The height of the lower layer influences the number of PEs, which is lower for 0.5~m and comparable for 0.75~m and 1~m, but does not affect the detection efficiency (plots not shown in this work). In all the configurations, the detection efficiency of the lower layer is underestimated in the same way due to geometrical constraints (see details about the calculation of the detection efficiency in Section~\ref{sec:analysis}). 
%
%
\subsection{Comparison of tank shapes}

The analysis described in the previous section was performed for the three tank shapes of interest: Circular-DLTs, Square-DLTs and Hexagonal-DLTs. In addition to tanks with circular base, which is the most commonly used shape for the construction of ground based astrophysical experiments due to their stronger strain resistance to the water pressure, ease of construction, and lower cost, tanks with square and the hexagonal bases have been considered because they would offer a higher fill factor, particularly important when it is necessary to cover areas with high density of tanks, like the inner array of the SWGO experiment. 

\begin{figure*}[h!]
	\centering
	\subfloat[]{\includegraphics[width=0.46\linewidth]{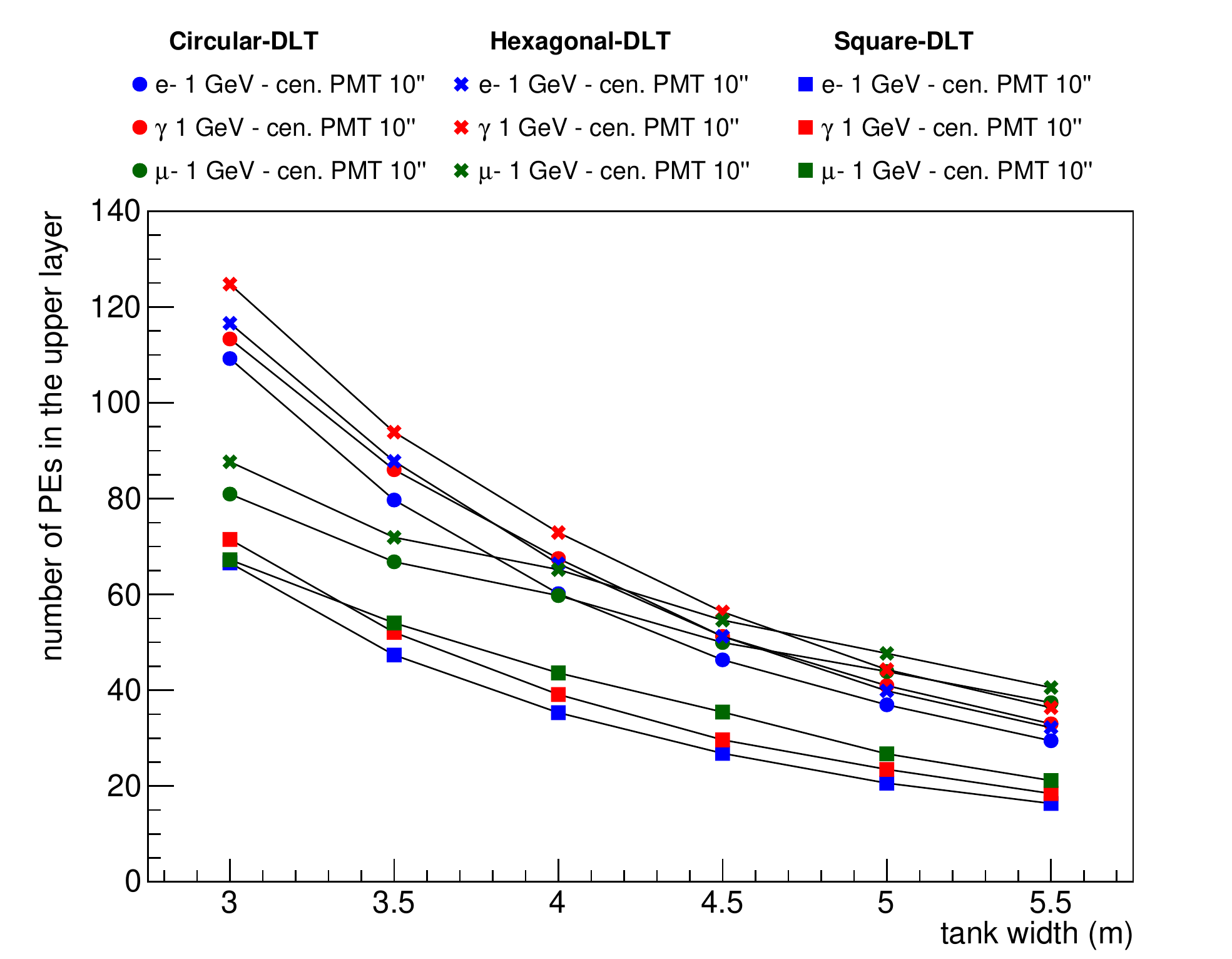}}
	\subfloat[]{\includegraphics[width=0.46\linewidth]{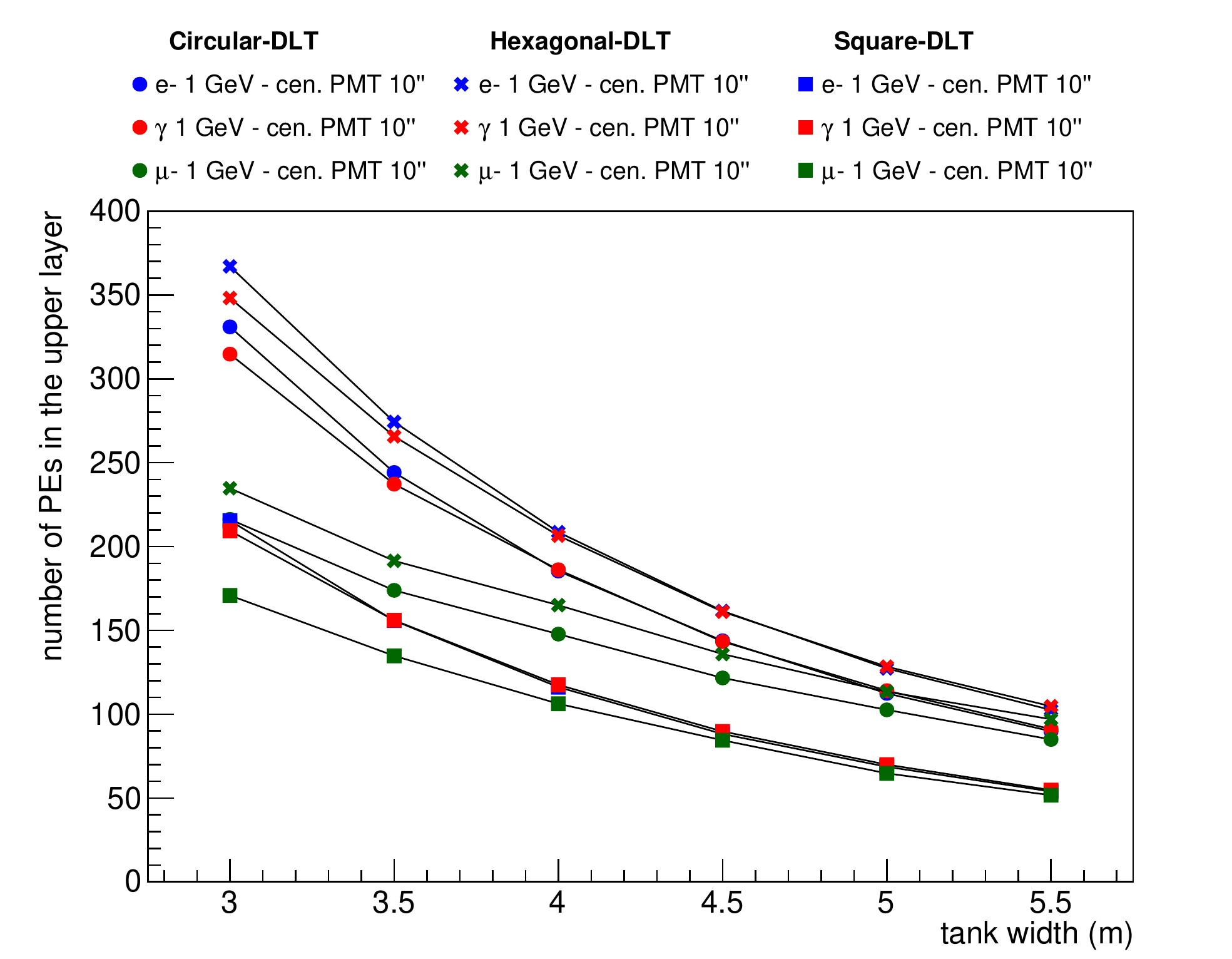}}
	\caption{Comparison of the number of PEs detected in the upper layer with non-reflective walls (a) and reflective walls (b), for 1~GeV particles, for different geometries and considering only the central 10\textquotedbl PMTs.} \label{fig:comp_npe_up_10in}
\end{figure*}

\begin{figure*}[h!]
	\centering
	\subfloat[]{\includegraphics[width=0.46\linewidth]{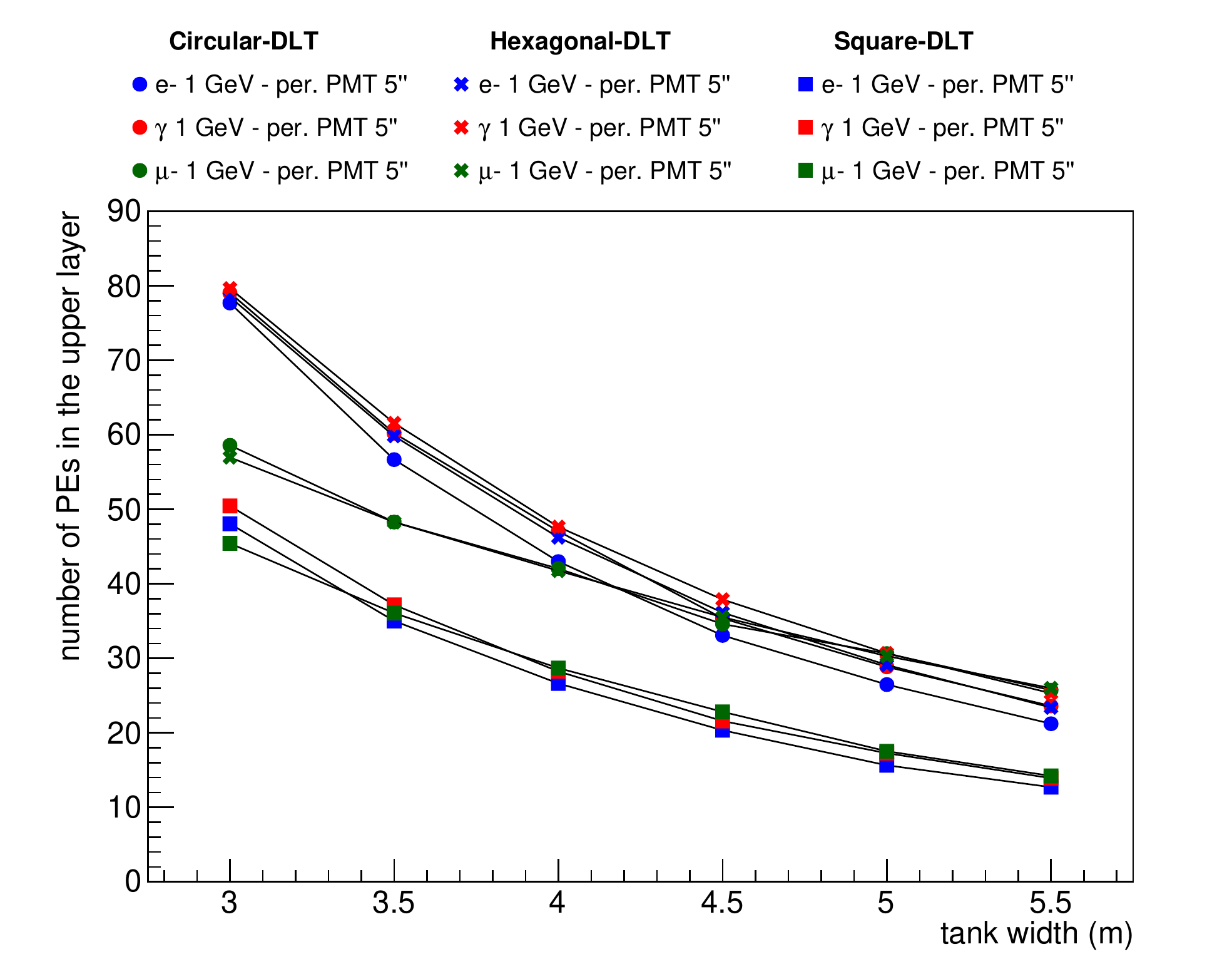}}
	\subfloat[]{\includegraphics[width=0.46\linewidth]{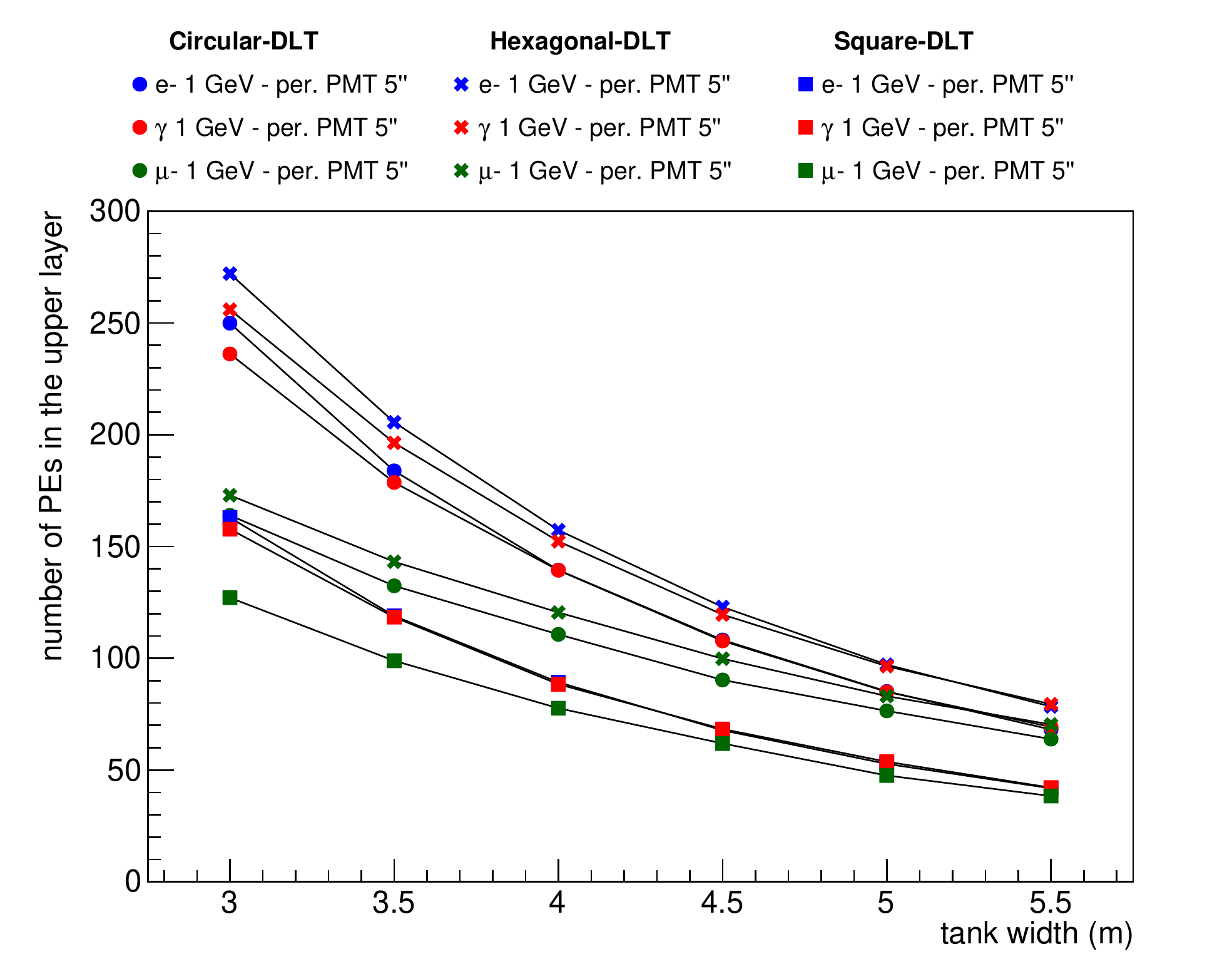}}
	\caption{Comparison of the number of PEs detected in the upper layer with non-reflective walls (a), and reflective walls (b) for 1~GeV particles, for different geometries and considering only the peripheral 5\textquotedbl PMTs.} \label{fig:comp_npe_up_5in}
\end{figure*}

\begin{figure*}[h!]
	\centering
	\subfloat[]{\includegraphics[width=0.46\linewidth]{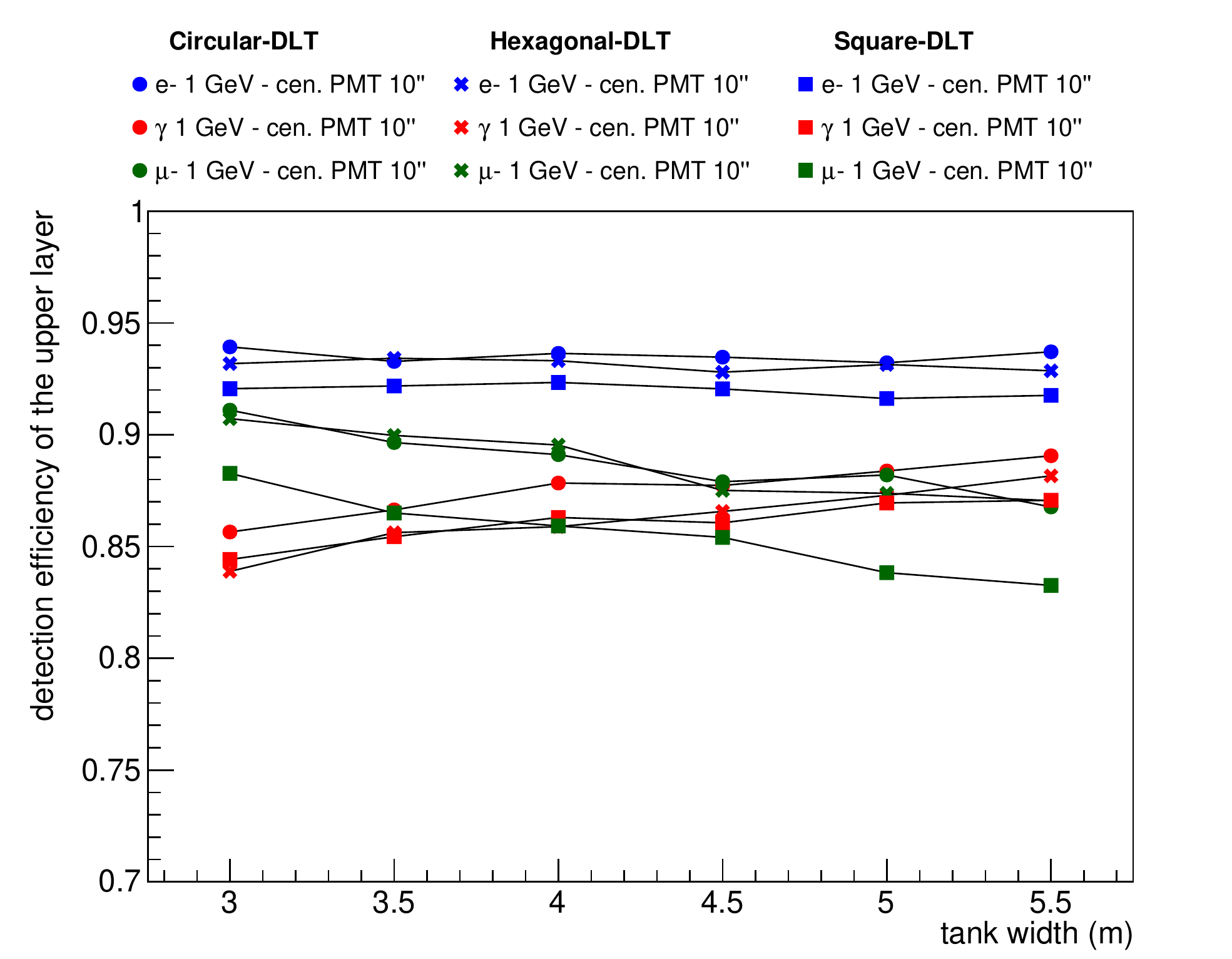}}
	\subfloat[]{\includegraphics[width=0.46\linewidth]{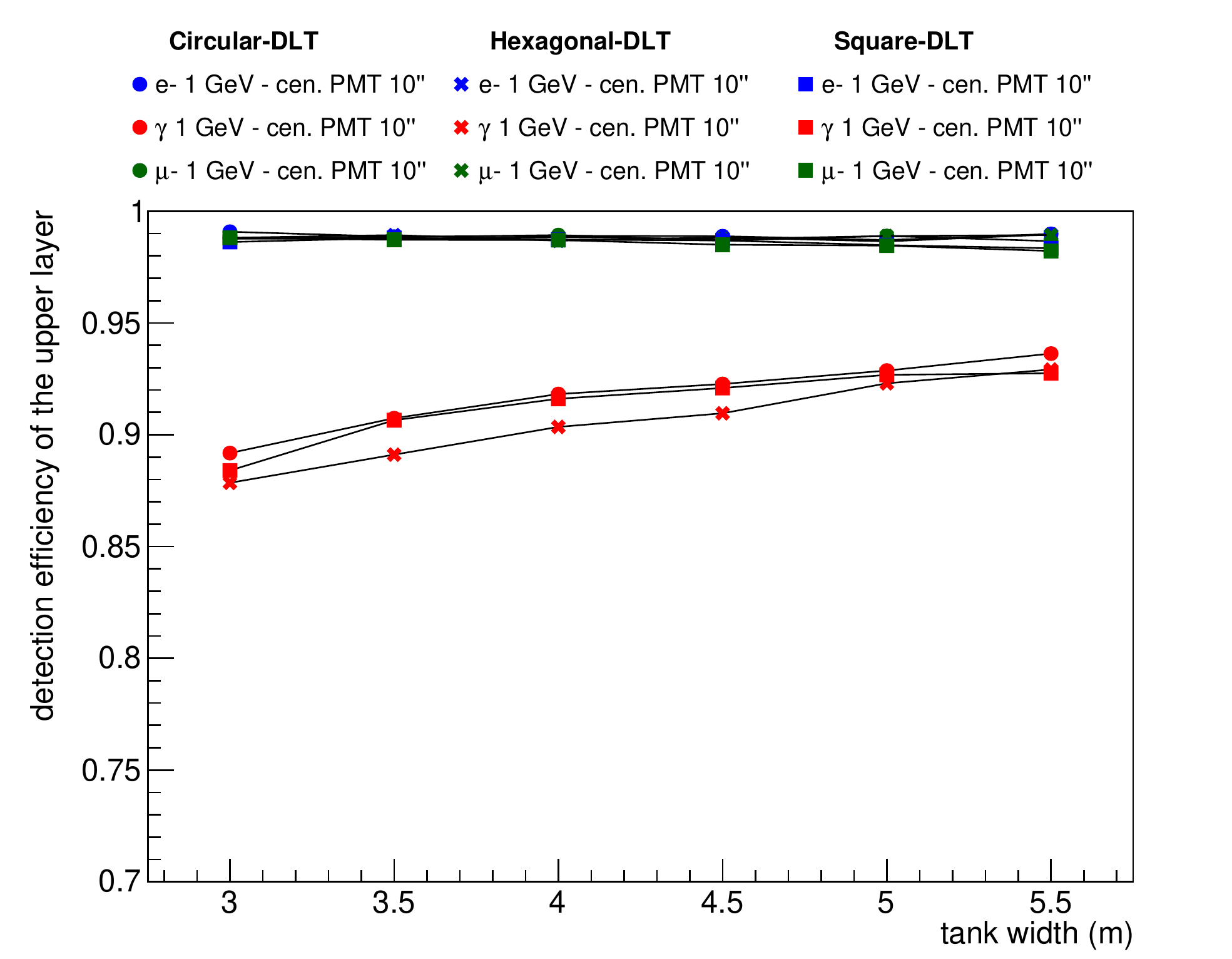}}
	\caption{Comparison of detection efficiency of the upper layer with non-reflective walls (a) and reflective walls (b), for 1~GeV particles, for different geometries and considering only the central 10\textquotedbl PMTs.} \label{fig:comp_eff_up_10in}
\end{figure*}

\begin{figure*}[h!]
	\centering
	\subfloat[]{\includegraphics[width=0.46\linewidth]{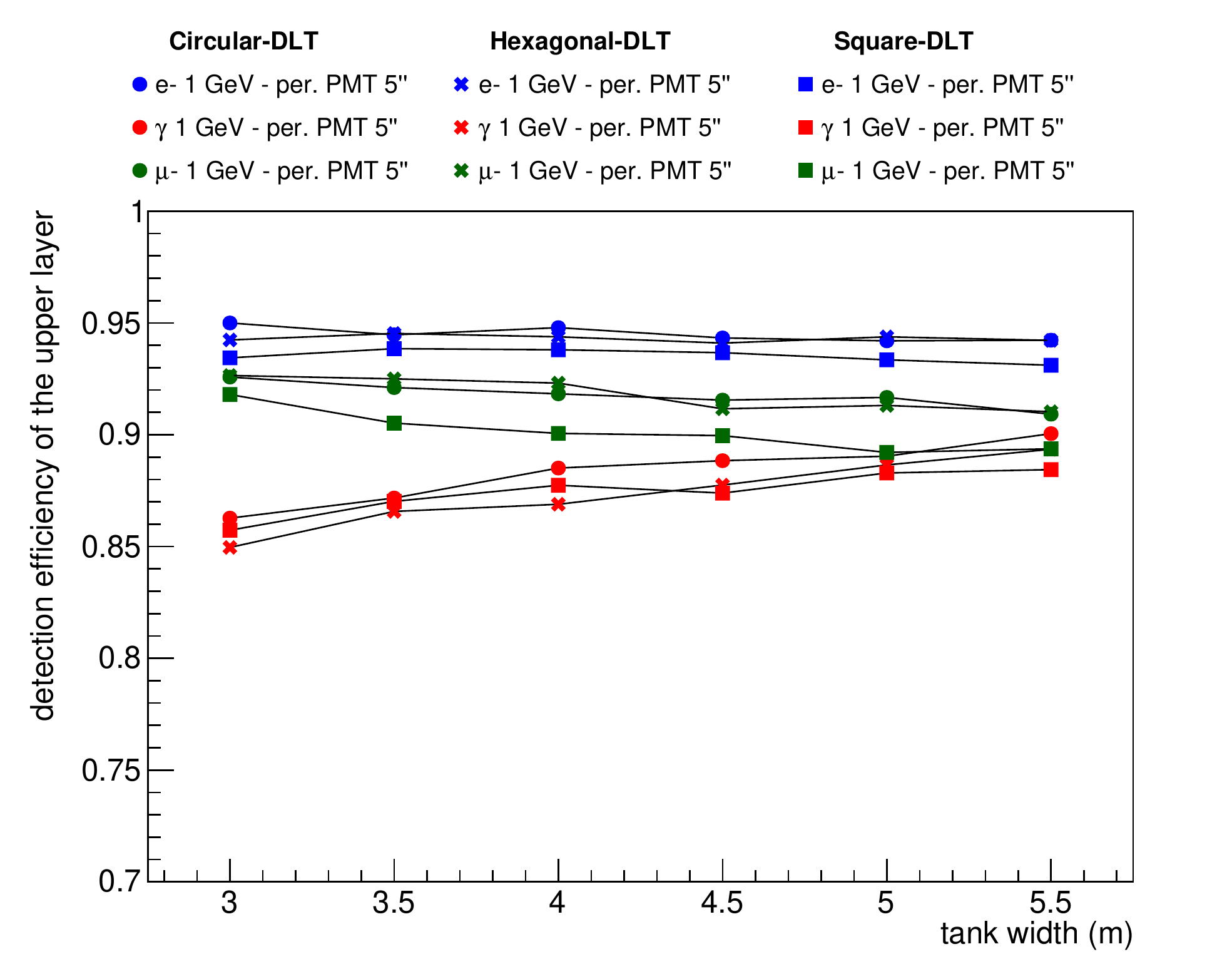}}
	\subfloat[]{\includegraphics[width=0.46\linewidth]{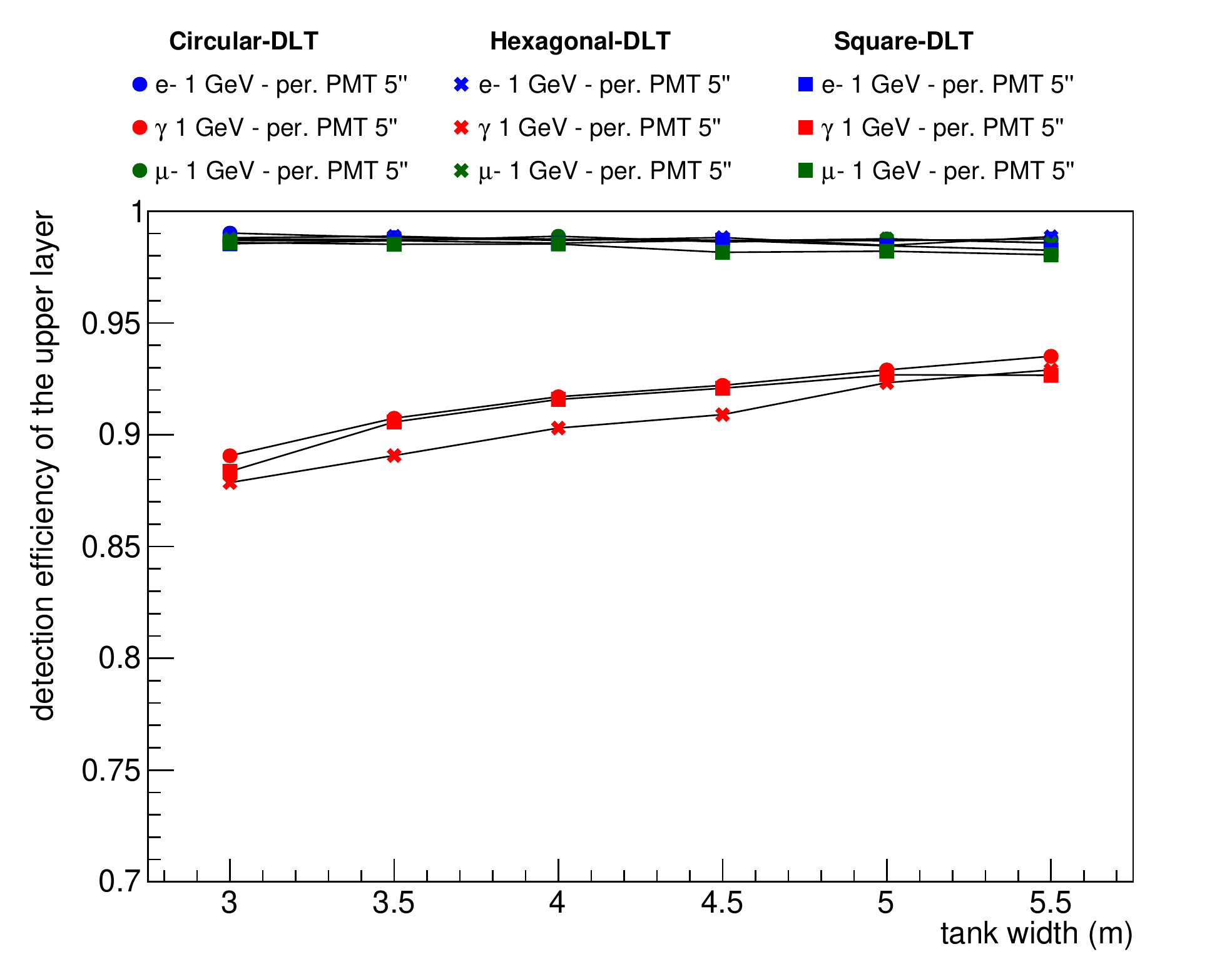}}
	\caption{Comparison of the detection efficiency of the upper layer with non-reflective walls (a) and reflective walls (b), for 1~GeV particles, for different geometries and considering only the peripheral 5\textquotedbl PMTs.} \label{fig:comp_eff_up_5in}
\end{figure*}

\begin{figure*}[h!]
	\centering
	\subfloat[]{\includegraphics[width=0.46\linewidth]{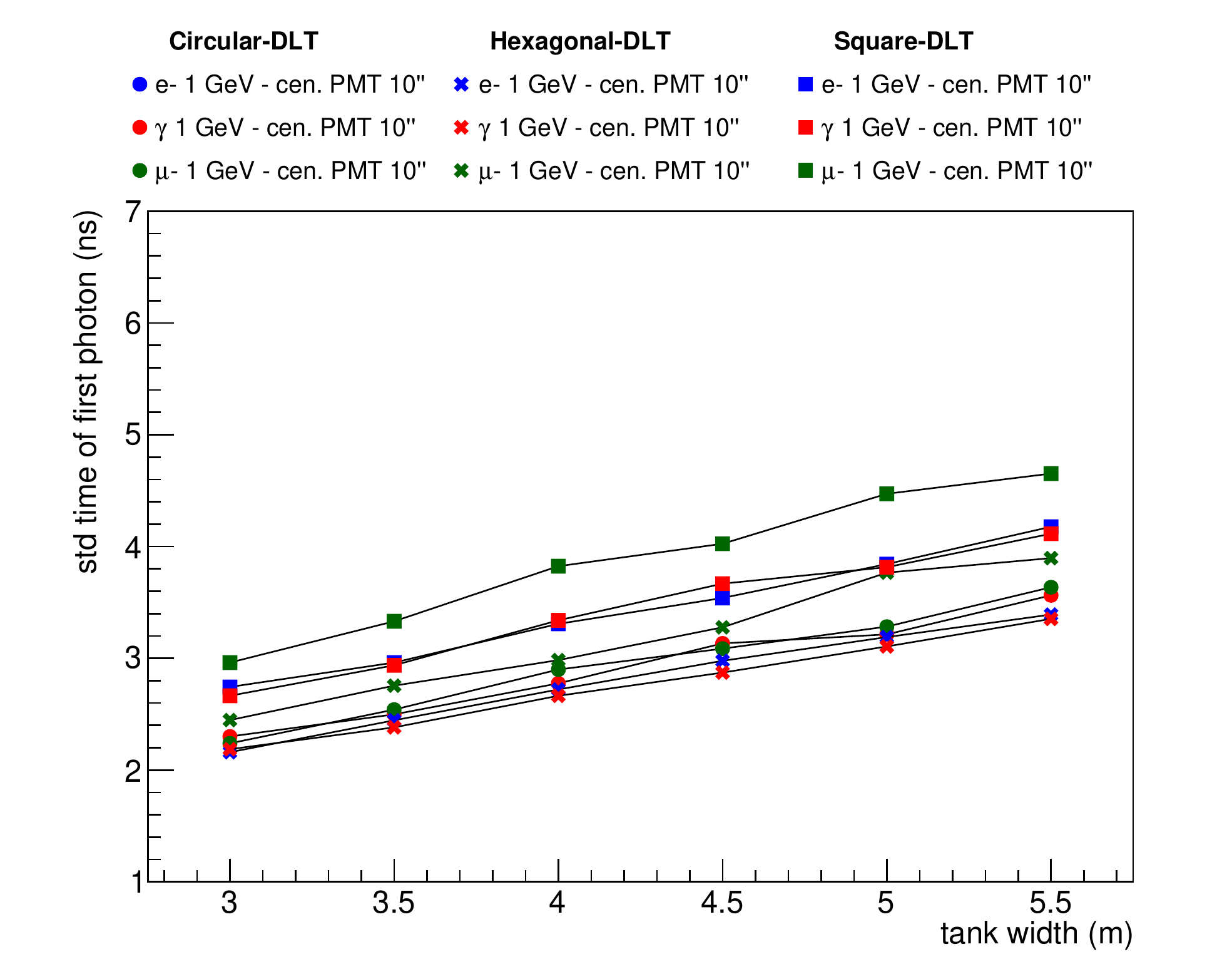}}
	\subfloat[]{\includegraphics[width=0.46\linewidth]{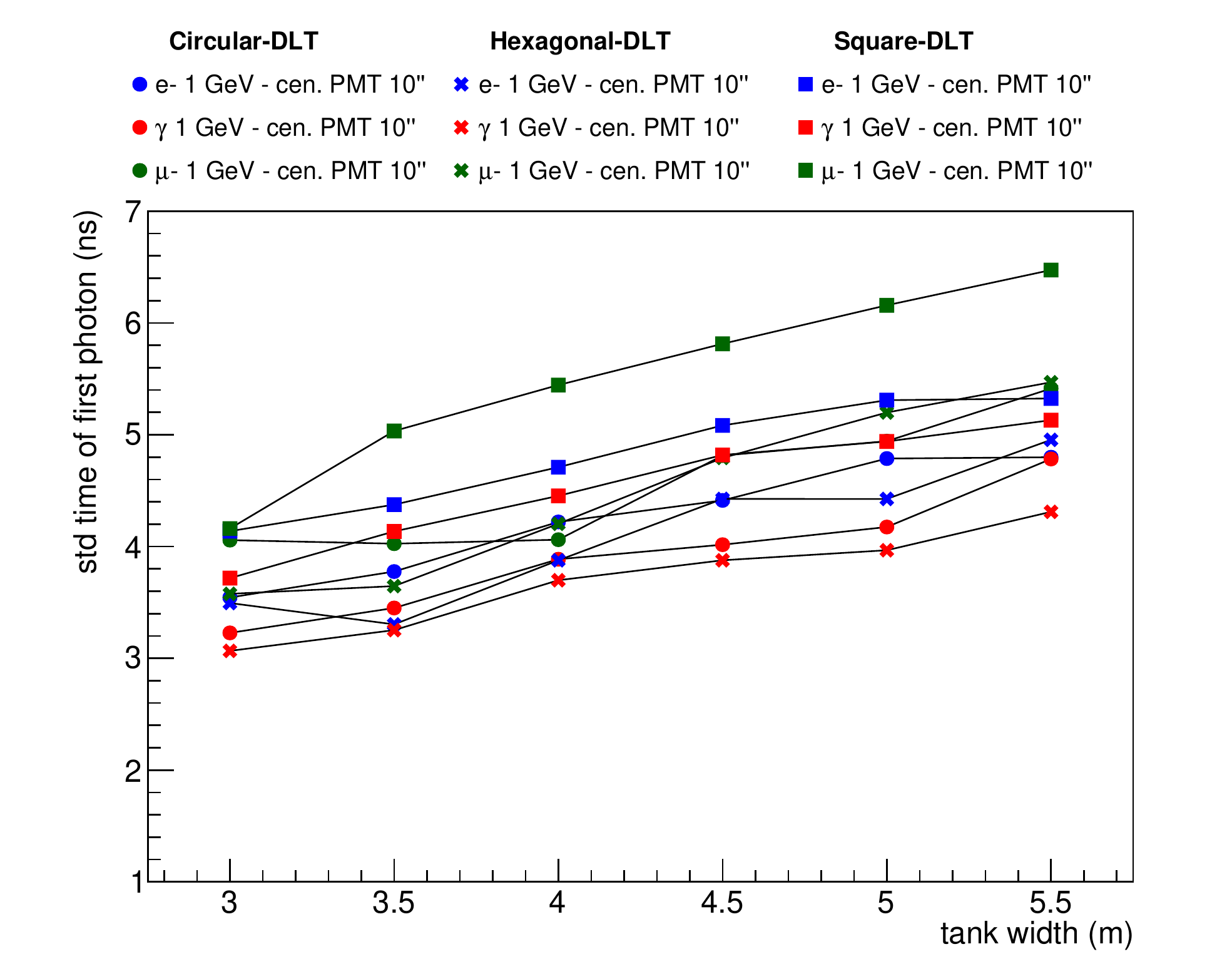}}
	\caption{Comparison of the time resolution of the measurement of the first photon in case of non-reflecting walls (a) and reflective walls (b) in the upper layer, for 1~GeV particles, for different geometries and considering only the peripheral 10\textquotedbl~PMTs in the upper layer.} \label{fig:comp_std_ftime_10in}
\end{figure*}

\begin{figure*}[h!]
	\centering
	\subfloat[]{\includegraphics[width=0.46\linewidth]{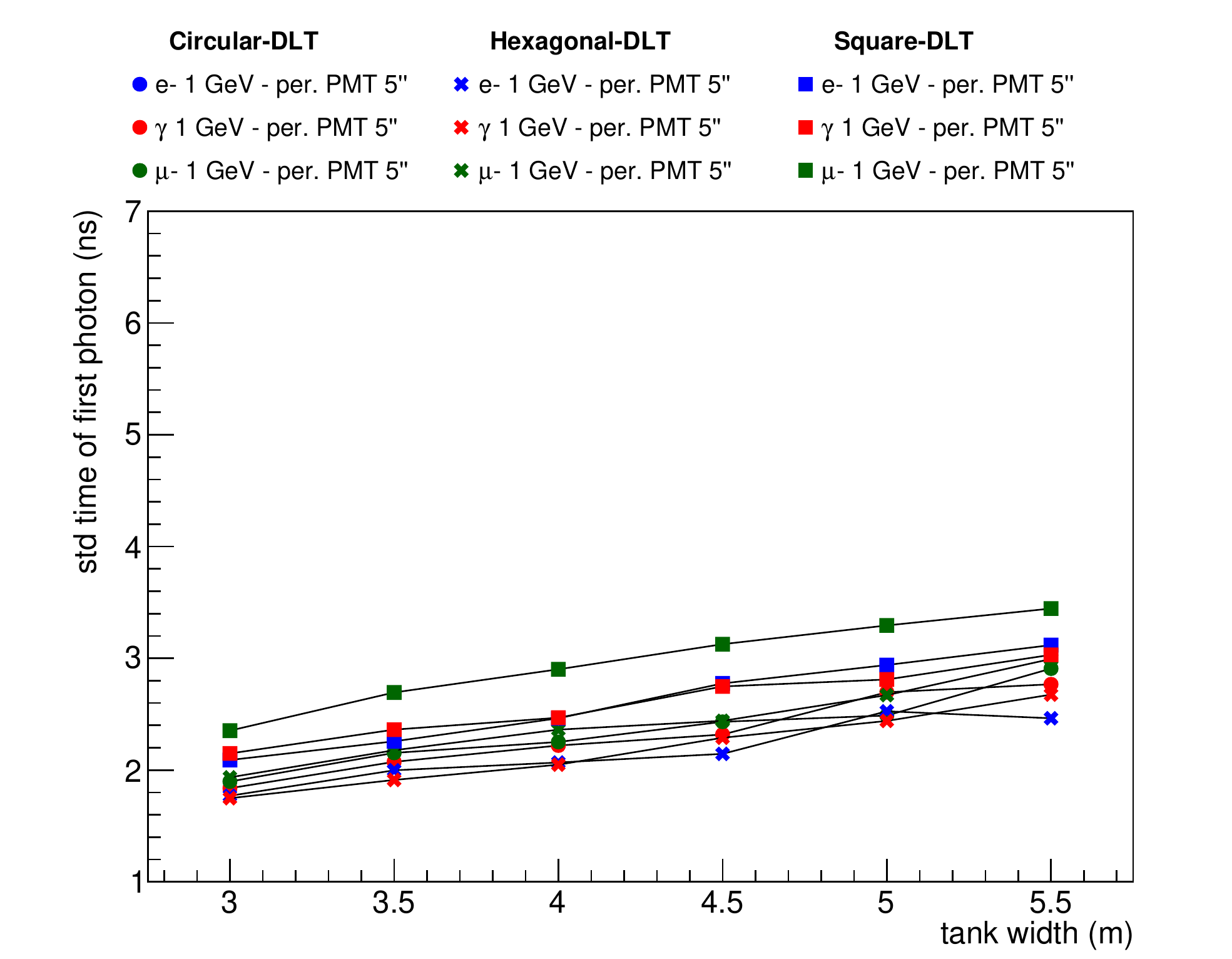}}
	\subfloat[]{\includegraphics[width=0.46\linewidth]{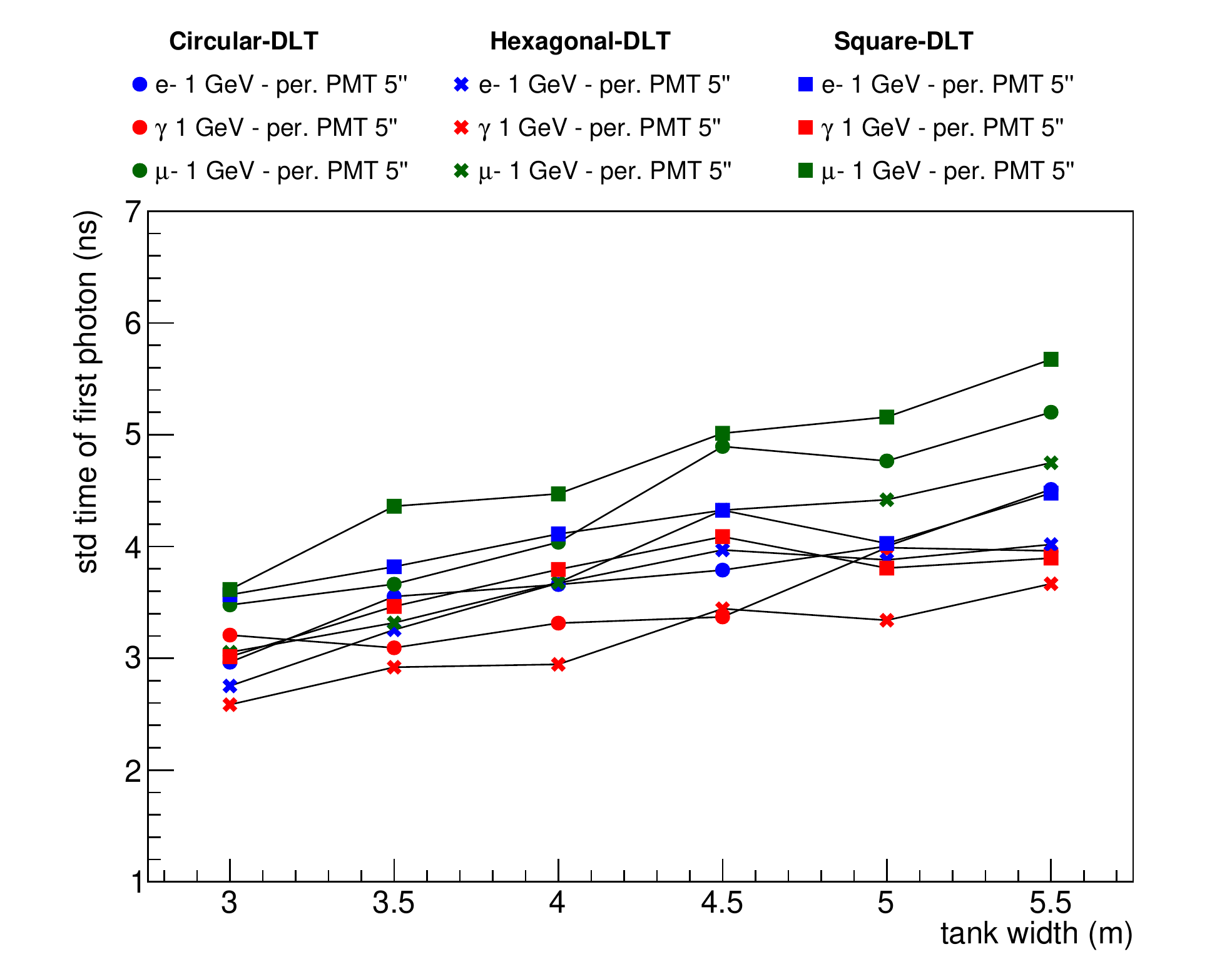}}
	\caption{Comparison of the time resolution of the measurement of the first photon in case of non-reflecting walls (a) and reflective walls (b) in the upper layer, for 1~GeV particles, for different geometries and considering only the peripheral 5\textquotedbl~PMTs in the upper layer.} \label{fig:comp_std_ftime_5in}
\end{figure*}

\begin{figure*}[h!]
	\centering
	\subfloat[]{\includegraphics[width=0.46\linewidth]{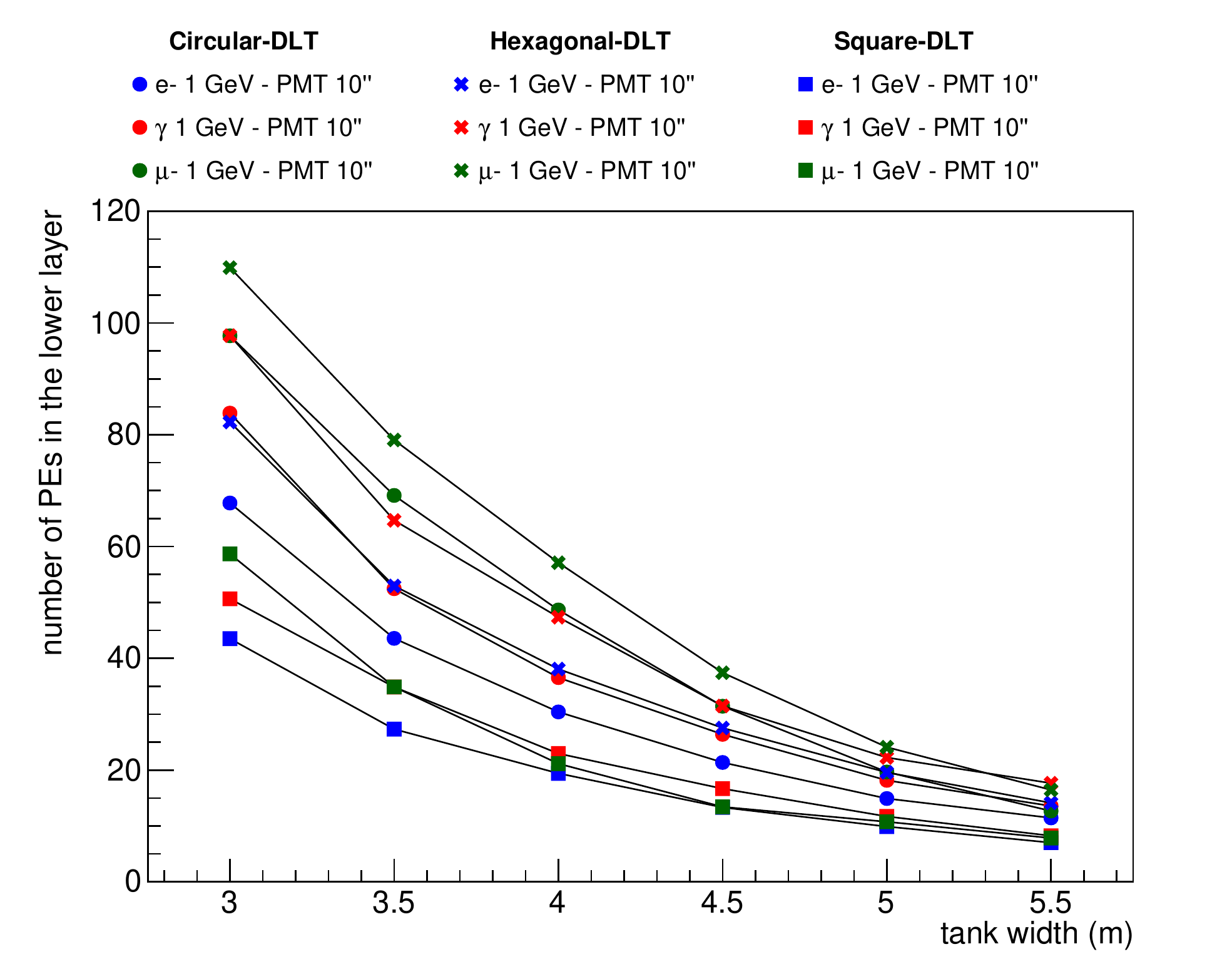}}
	\caption{Comparison of the number of PEs in the lower layer with reflective walls, for 1~GeV particles, for different geometries and considering the 10\textquotedbl~PMT.} \label{fig:comp_npe_low_10in}
\end{figure*}

\begin{figure*}[h!]
	\centering
	\subfloat[]{\includegraphics[width=0.46\linewidth]{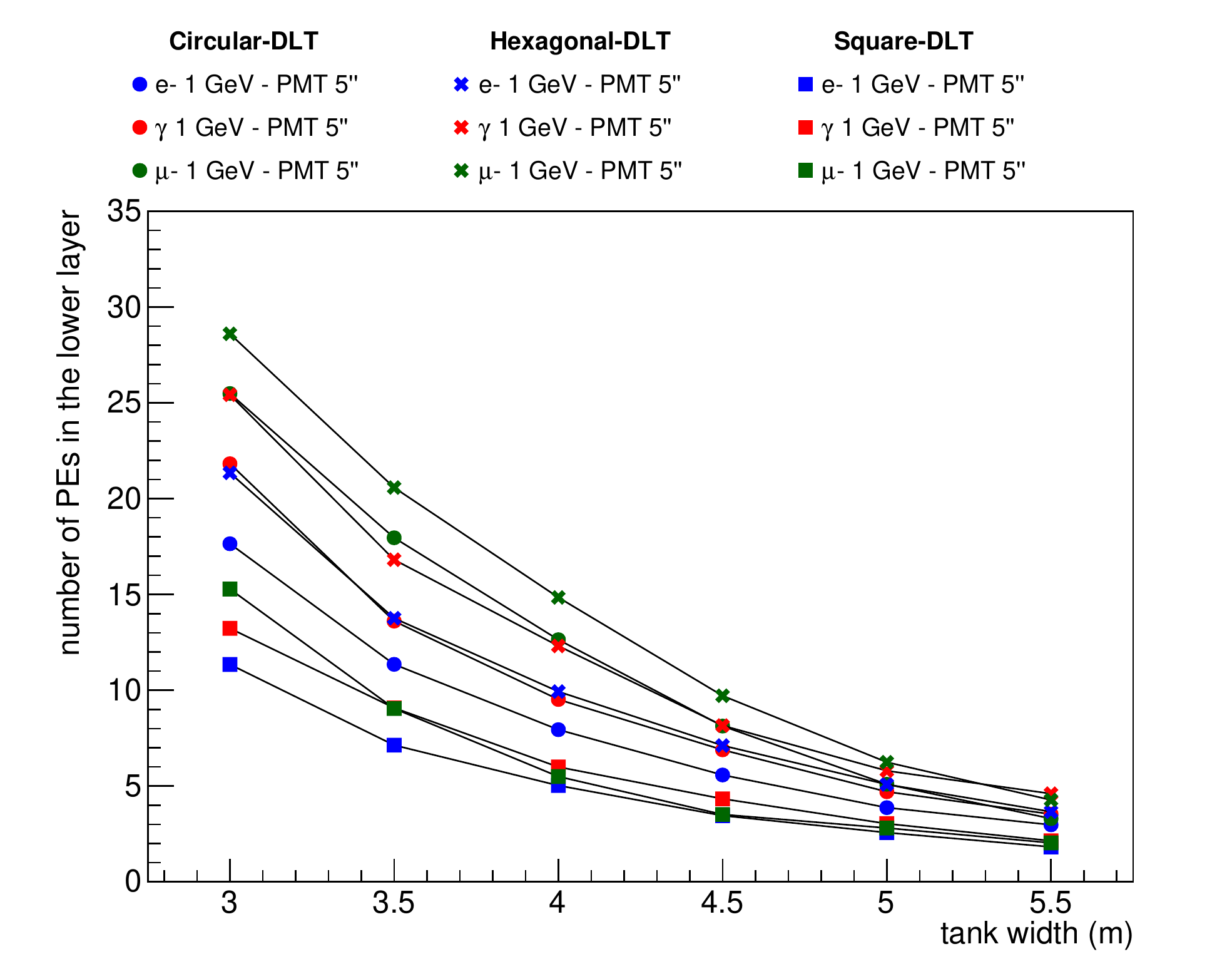}}
	\caption{Comparison of the number of PEs in the lower layer with reflective walls, for 1~GeV particles, for different geometries and considering the 5\textquotedbl~PMT.} \label{fig:comp_npe_low_5in}
\end{figure*}

\begin{figure*}[h!]
	\centering
	\subfloat[]{\includegraphics[width=0.46\linewidth]{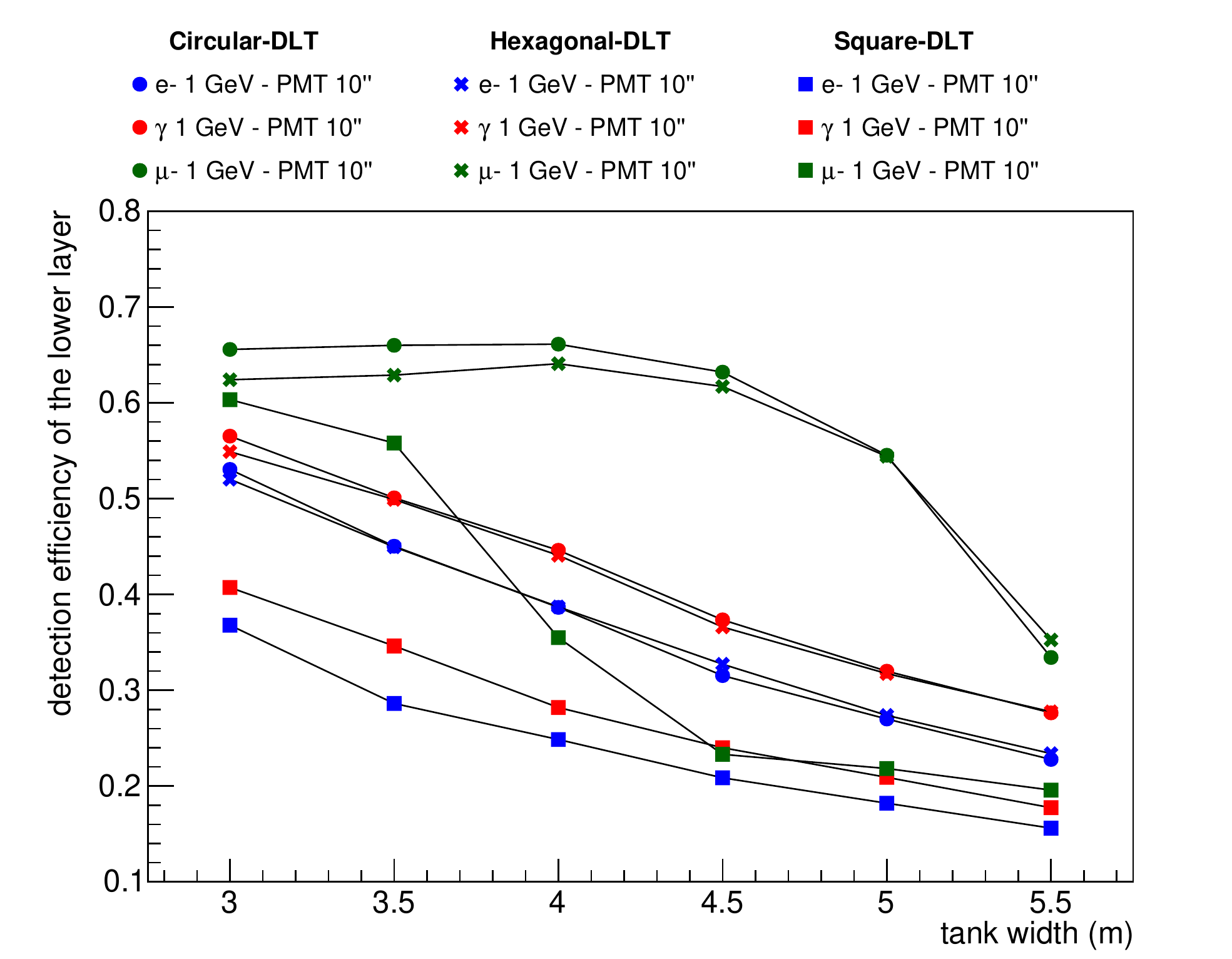}}
	\caption{Comparison of the detection efficiency of the lower layer with reflective walls, for 1~GeV particles, for different geometries and considering the 10\textquotedbl~PMT. In all the configurations, it is underestimated due to geometrical constraints.} \label{fig:comp_eff_low_10in}
\end{figure*}

\begin{figure*}[h!]
	\centering
	\subfloat[]{\includegraphics[width=0.46\linewidth]{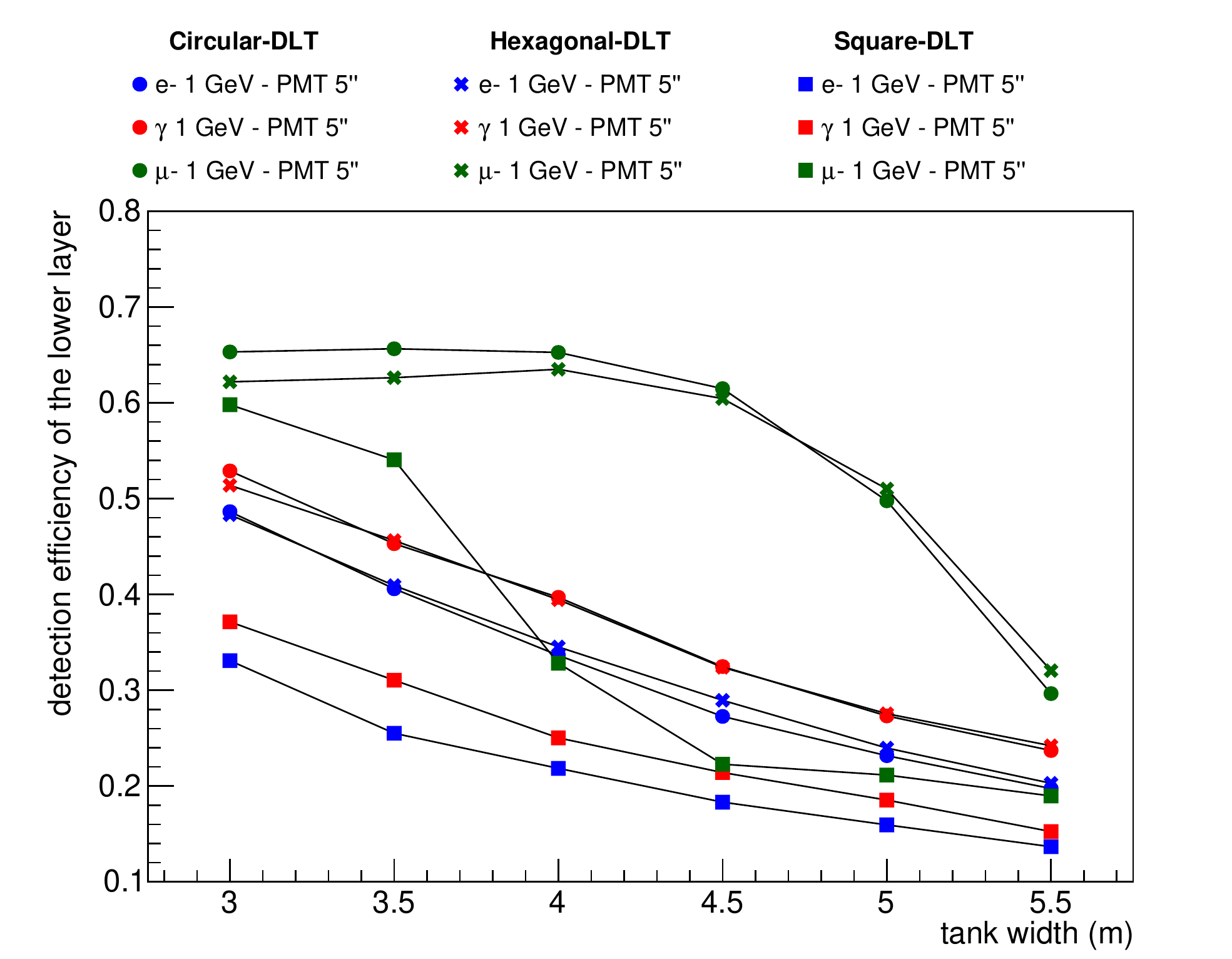}}
	\caption{Comparison of the detection efficiency of the lower layer with reflective walls, for 1~GeV particles, for different geometries and considering the 5\textquotedbl~PMT. In all the configurations, it is underestimated due to geometrical constraints.} \label{fig:comp_eff_low_5in}
\end{figure*}

The plots shown in this section represent the aforementioned parameters in function of the size of the tanks with different geometries and were made for 1~GeV particles, in order to compare the tank performance in the same conditions. The size of the tanks is represented by their width, i.e. diameter for Circular-DLT, two times the side for Hexagonal-DLT, and side for Square-DLT.
In Fig.~\ref{fig:comp_npe_up_10in}--\ref{fig:comp_std_ftime_5in} the performance of the upper layers of tanks with different shapes are shown, considering a central 10\textquotedbl~PMT or four peripheral 5\textquotedbl~PMTs. Panels (a) are relative to non-reflecting walls, while panels (b) are for reflecting walls. Similarly, in Fig.~\ref{fig:comp_npe_low_10in}--\ref{fig:comp_eff_low_5in} the performance of the lower layer is shown considering a 10\textquotedbl~PMT or a 5\textquotedbl~PMT. 

In general the Circular-DLT and Hexagonal-DLT produce more PEs than the Square-DLT.
%
%
With reflective walls, the number of PEs is about three times that obtained with non-reflective walls, in any kind of tank (compare panels (a) with panels (b) in Fig.~\ref{fig:comp_npe_up_10in} and Fig.~\ref{fig:comp_npe_up_5in}).  
Comparing the response of the central 10\textquotedbl~PMT and the four peripheral 5\textquotedbl~PMTs in the upper layer, the number of PEs for the first configuration is higher than the number of PEs for the latter configuration for all tank geometries (compare panel (a) and panel (b) in Fig.~\ref{fig:comp_npe_up_10in} with the same panels in Fig.~\ref{fig:comp_npe_up_5in}). 
%
The detection efficiency of the upper layer is similar for Circular-DLTs and Hexagonal-DLTs, while it is slightly lower for Square-DLTs. As already shown, it is higher using reflective walls (compare panels (a) with panels (b) in Fig.~\ref{fig:comp_eff_up_10in} and Fig.~\ref{fig:comp_eff_up_5in}). 
Although the number of PEs is higher for the central 10\textquotedbl~PMT, the detection efficiency for particles of 1~GeV is similar for both configurations of PMTs (compare panels (a) and (b) in Fig.~\ref{fig:comp_eff_up_10in} with those in Fig.~\ref{fig:comp_eff_up_5in}).
%
%
%
The temporal resolution of the measurement of the first photon considering the central 10\textquotedbl~PMT is $\sim$0.5~ns larger than that of the four peripheral 5\textquotedbl~PMTs (compare panels (a) and panels (b) in Fig.~\ref{fig:comp_std_ftime_10in} with those in Fig.~\ref{fig:comp_std_ftime_5in}). However, such values are slightly higher for Square-DLT than the others.
In general, with reflective walls the temporal resolution of the measurement of the first photon in upper layers is about $1-2$~ns larger than with non-reflective walls, in all kinds of tanks (compare panels (a) with panels (b) in Fig.~\ref{fig:comp_std_ftime_10in} and Fig.~\ref{fig:comp_std_ftime_5in}). 
%
%

In the lower layer of the three kinds of tanks, the number of PEs produced by the 10\textquotedbl~PMT is four times that produced by the 5\textquotedbl~PMT, simply because of the ratio between the active areas of the sensors (see Fig.~\ref{fig:comp_npe_low_10in} and Fig.~\ref{fig:comp_npe_low_5in}). The height of the lower layer influences the number of PEs, which is lower for 0.5~m but comparable for 0.75~m and 1~m.
%

Despite the difference in the number of PEs detected with the two PMT configurations, the detection efficiency of the lower layer does not vary (see Fig.~\ref{fig:comp_eff_low_10in} and Fig.~\ref{fig:comp_eff_low_5in}). It is similar for Circular-DLT and Hexagonal-DLT and higher than for Square-DLT, and it is similar also for the three different heights considered. In all the configurations, the detection efficiency of the lower layer is underestimated in the same way due to geometrical constraints (see details about the calculation of the detection efficiency in Section~\ref{sec:analysis}).
%
%

\clearpage 

\section{Conclusion}

This study allowed to compare the response of double-layer tanks with different shapes, i.e. with circular, hexagonal and square base of different size, to the passage of single particles of different type and energy. Moreover, it offered the possibility to compare tanks with reflective and non-reflective walls in the upper layer.

We found that regardless of the tank design and the reflective properties of the walls in the upper layer, the performance of the tanks worsen while increasing the width of the tank, because the ``sensitive area'', i.e. the area covered by the PMTs, decreases with respect to that of the base of the tank.

By using reflective walls instead of non-reflective walls in the upper layers, the detection efficiency increases, but the time resolution of the measurement of the first photon widen, in particular for particle with low energy.

In lower layers, electrons and gamma-rays of 10~MeV and 100~MeV are rarely detected. The height of the lower layer influences the number of PEs, which is lower for 0.5~m but comparable for 0.75~m and 1~m. Howewer, the detection efficiency of the lower layer does not vary much with its height.

Comparing the performance of the central 10\textquotedbl~PMT and the four peripheral 5\textquotedbl~PMTs in the upper layer, the first configuration produces more PEs than the second one, although both sensitive areas are similar, but the detection efficiencies are comparable. In a similar comparison for the lower layer, the 10\textquotedbl~PMT produces more PEs than the 5\textquotedbl~PMT due to the larger area of the photocathode, but the detection efficiencies are similar.

The comparison of the performance of tanks with different geometries revealed that the Circular-DLTs and Hexagonal-DLTs have similar performance, which is better than that of the Square-DLTs.
Nevertheless, for the final design of ground based astrophysical observatories like the SWGO array, it should be taken into account also that with Hexagonal-DLT and Square-DLT it is possible to achieve a higher fill factor, although they are potentially more expensive solutions.

\acknowledgments

We thank our colleagues within the SWGO Collaboration for the discussions and the software framework used in this work. We thank the
HAWC Collaboration for providing the AERIE software.


\end{document}